\documentclass[showpacs,preprintnumbers,secnumarabic,amsmath,amssymb,nofootinbib]{revtex4}
\usepackage{graphics}
\usepackage[export]{adjustbox}
\usepackage{amsmath}

\pdfoutput=1

\usepackage{tcolorbox}
\graphicspath{{},{Logos/}}
\usepackage{graphicx,color}

\usepackage{bm}
\usepackage{amsfonts}
\usepackage{amsmath}
\usepackage{amssymb}
\usepackage{float}
\usepackage{hyperref}
\usepackage{subfigure}
\usepackage{slashed}
\usepackage{tikz}
\usetikzlibrary{arrows,shapes}
\usetikzlibrary{trees}
\usetikzlibrary{matrix,arrows} 				
\usetikzlibrary{positioning}				
\usetikzlibrary{calc,through}				
\usetikzlibrary{decorations.pathreplacing}  
\usepackage{pgffor}							

\usetikzlibrary{decorations.pathmorphing}	
\usetikzlibrary{decorations.markings}
\tikzset{
	vector/.style={decorate, decoration={snake}, draw},
	provector/.style={decorate, decoration={snake,amplitude=2.5pt}, draw},
	antivector/.style={decorate, decoration={snake,amplitude=-2.5pt}, draw},
	fermion/.style={draw=black, postaction={decorate},
		decoration={markings,mark=at position .55 with {\arrow[draw=black]{>}}}},
	fermionbar/.style={draw=black, postaction={decorate},
		decoration={markings,mark=at position .55 with {\arrow[draw=black]{<}}}},
	fermionnoarrow/.style={draw=black},
	gluon/.style={decorate, draw=black,
		decoration={coil,amplitude=4pt, segment length=5pt}},
	scalar/.style={dashed,draw=black, postaction={decorate},
		decoration={markings,mark=at position .55 with {\arrow[draw=black]{>}}}},
	scalarbar/.style={dashed,draw=black, postaction={decorate},
		decoration={markings,mark=at position .55 with {\arrow[draw=black]{<}}}},
	scalarnoarrow/.style={dashed,draw=black},
	electron/.style={draw=black, postaction={decorate},
		decoration={markings,mark=at position .55 with {\arrow[draw=black]{>}}}},
	bigvector/.style={decorate, decoration={snake,amplitude=4pt}, draw},
}

\tikzstyle{block} = [draw, rectangle, 
minimum height=3em, minimum width=6em]

\newcommand{\be}{\begin{equation}}
\newcommand{\ee}{\end{equation}}
\newcommand{\bea}{\begin{eqnarray}}
\newcommand{\eea}{\end{eqnarray}}

\newcommand{\gev}{{\rm GeV}}

\begin{document}
	
\title{Gravity Waves and Primordial Black Holes in Scalar Warm Little Inflation}

\author{Mar Bastero-Gil} \email{mbg@ugr.es} \affiliation{Departamento
  de F\'{\i}sica Te\'orica y del Cosmos, Universidad de Granada,
  Granada-18071, Spain}

\author{Marta Sub\'{\i}as Díaz-Blanco} \email{subias.marta@gmail.com} \affiliation{Departamento
  de F\'{\i}sica Te\'orica y del Cosmos, Universidad de Granada,
  Granada-18071, Spain}

\begin{abstract}
 In warm inflation, dissipation  due to the interactions of the inflaton field to other light degrees of freedom leads naturally to the enhancement of the primordial spectrum during the last 10-20 efolds of inflation. We study this effect in a variant of the Warm Little Inflaton model, where the inflaton couples to light scalars, with a quartic chaotic potential. These large fluctuations on re-entry will form light, evaporating Primordial Black Holes, with masses lighter than $10^6$ g. But at the same time they will act as a source for the tensors at second order. The enhancement is maximal near the end of inflation, which result in a spectral density of Gravitational Waves (GW) peaked at frequencies $f \sim O(10^5-10^6)$ Hz today, and with an amplitude $\Omega_{GW} \sim 10^{-9}$. Although the frequency range is outside the reach of present and planned GW detectors, it might be reached in future high-frequency gravitational waves detectors, designed to search for cosmological stochastic GW backgrounds above MHz frequencies.

\end{abstract}



\maketitle


\section{Introduction}
Inflation, a period of accelerated expansion in the early evolution of the Universe, provides an elegant solution to the horizon and flatness problem, and at the same time a mechanism to generate the primordial seeds required to explain the observed large scale structure \cite{inflation1, inflation2, inflation3}. Such a period can be easily modelled by a dynamical scalar field, the inflaton, with the appropriate potential and interactions. According to the most recent CMB data \cite{Planckinf}, a successful inflationary model should predict a quasi-adiabatic and gaussian primordial spectrum of perturbations, with spectral index $n_s=0.965 \pm 0.004$, and less than a 10\% of primordial gravity waves (GW), i.e, a tensor-to-scalar ratio below $r < 0.07$. However, although inflation should last at least around 50-60 efolds in order to explain the horizon and flatness problem,  CMB constraints only apply to roughly the first 10 e-folds of those, when the larger observable scales $k \simeq 10^{-3}-1$ Mpc$^{-1}$ leave the horizon, leaving the remaining inflationary dynamics and the primordial spectrum largely unconstrained. At smaller scales, the spectral index can turn from red-tilted to blue-tilted, and the amplitude of the primordial spectrum can be much larger than the CMB value $P_{\cal R} = 2.1\times 10^{-9}$. If the amplitude reaches a critical value $P_{\cal R} \sim 10^{-2}$ this could lead to the formation of Primordial Black Holes (PBHs) on re-entry, due to the collapse of the overdensities \cite{PBH1,  PBHGreen, PBHKhlopov,PBHSasaki}, with a very rich phenomenology. For example non-evaporating PBHs within the appropriate mass range $M_{PBH} > 10^{15}$ g could be all or part of the Dark Matter content of our Universe \cite{PBHDMClesse, PBHDMCarr, PBHDMJuan, Pi:2017gih, PBHDMBallesteros}; and they can also act as a source of GW with a characteristic frequency spectrum \cite{PBHGWNakama, PBHGWJuan,  Cai:2018dig, PBHGWSukannya}. Similarly, large scalar perturbations are a source of tensor perturbations at second order, and therefore the same mechanism that leads to PBHs during inflation will lead to a larger amplitude of primordial GW on smaller scales \cite{GWMollerach, GWWands, GWBaumann}. 

While the study of PBHs formation and the generation of primordial GW has been actively pursued over recent years in different inflationary models, there has been not so many studies in the context of warm inflation. In standard, cold inflationary scenarios, inflaton interactions play no role neither in the slow-roll dynamics not in the generation of the primordial spectrum; on the contrary in warm inflation, those interactions can lead to the partial dissipation of the inflaton energy density into radiation already during inflation  \cite{warmstandard, warmstandard2, warmstandard3}. The presence of a subdominant thermal bath  can modify the dynamics both at the background and the perturbation level. When dissipation dominates over the Hubble friction term, the motion of the inflaton field will be further slow-down, enlarging the duration of inflation. On one hand slow-roll conditions are easier to fulfil, and in particular the inflaton mass can be closer to the Hubble parameter during inflation, relaxing the so-called ``$\eta$''-problem \cite{etaproblem,etaproblem2}. And in single-field models of inflation like chaotic-like ones, the last 60 efolds of inflation can take place at smaller values of the inflaton energy density and the Hubble parameter $H$, meaning a smaller value of the primordial tensor perturbations and the tensor-to-scalar ratio. Most importantly, the presence of the thermal bath means that inflaton perturbations acquire now a thermal component on top of the standard vacuum one, with implications also for the properties of the scalar primordial spectrum. Models that are excluded by observations in their simpler cold inflation version, like quartic and quadratic chaotic models, are perfectly compatible with Planck data in their warm version \cite{chaoticwarm}.

Dissipation will depend on the temperature of the thermal bath, and although during inflation everything takes place in the slow-roll, slowly changing regime, what matters is the comparison between the dissipation coefficient $\Upsilon$ and Hubble friction $H$, given by the ratio $Q=\Upsilon/3H$. Typically this ratio is an increasing function of time, and depending on parameters inflation may enter in what is called the strong dissipative regime (SDR), i.e., $Q \gg 1$ with dissipation fully dominating the dynamics at the background and perturbative level. Thermal fluctuations must be taken into account \cite{growing1}, and  when the dissipative coefficient is an increasing function of  the temperature $T$ this leads to the amplification of the scalar curvature perturbation,  and eventually to a blue-tilted spectrum. While we may want to avoid as much as possible this effect on CMB scales, later on before the end of inflation this effect may enhance the spectrum enough to source the tensor at second order and allow the formation of PBHs. The latter has been studied (to our knowledge for the first time) in Ref. \cite{Arya} for a dissipative coefficient $\Upsilon \propto T^3$ and an inflationary quartic chaotic model. Sufficient enhancement of the primordial spectrum only takes place during the last 10-15 e-folds of inflation, which translates at re-entry in relatively light, evaporating PBHs with masses $M_{PBH} \lesssim 10^3$ g. The effect on the tensor spectrum has been studied in  Ref. \cite{tensorswarm2} for a warm quintessential model with a linear and cubic dissipative coefficient. They focused on the enhancement during the kinetion period following warm quintessential inflation. The GW spectrum today would have a large enough amplitude to be detected,  but peaked at too large frequencies,  by far outside the range of present and future GW detectors. However this conclusion may depend on the functional $T$ dependence of the dissipative coefficient, i.e., the pattern of inflaton interactions, and to some extend on the inflationary model. Here we want to extend their analyses and explore other possibilities. 

A cubic dissipative coefficient results for example when the inflaton couples to the light degrees of freedom (dof) through a heavy mediator. This pattern directly protects the inflaton potential from acquiring large thermal corrections that might spoil inflation, but typically requires a too large no. of light and mediator fields in order to get enough dissipation. Instead the inflaton potential can be shielded against large radiative corrections by the use of symmetries. When coupled directly to fermionic light dof this leads to a linear dissipative coefficient in the so-called ``Warm Little Inflation'' (WLI) model \cite{WLI, WLImodels, joaoluis}; an axion-like coupling to Yang-Mills fields gives again $\Upsilon \sim T^3$ in the ``Minimal Warm Inflation'' (MWI) model \cite{MWI}; while replacing fermionic by light scalar dof in the WLI give rise to and inverse dissipative coefficient $\Upsilon \sim 1/T$ \cite{SWLI}. In the latter case, we have ``light'' particles when their masses $m_i$ are below $T$, but as inflation proceeds and $T$ decreases, by the end of inflation we may have $m_i \gg T$, and $\Upsilon \sim T^\kappa$ with $\kappa>0$. While an inverse dissipative coefficient is free of the problems of the growing mode at CMB scales, before the end of inflation we recover the enhancement of the spectrum and the prospects to have PBHs and non-negligible GWs. This is the pattern that we want to study in this paper.

For that we need to derive the analytical expression of the scalar primordial spectrum beyond CMB scales, i.e., the expression of the ``growing mode''. This would require the numerical integration of stochastic equations \cite{warmstandard3, growing1}, for different values of the model parameters (couplings and masses). However, the amplitude and tilt of the scalar spectrum mainly depend on the value of $\kappa=d \ln \Upsilon/d \ln T$ when the fluctuation leaves the horizon. Therefore instead of scanning over the model parameters, we will first derive the expression of the growing mode with $\Upsilon \propto T^\kappa$, for different values of constant $\kappa$. Now we can scan the model parameters, derive the value of $\kappa$, and then the spectrum at different stages during inflation: constraints on CMB scales set the parameter space consistent with observations, and for that we can explore whether it leads or not to enough amplification of GW, and PBHs. We will work with a quartic chaotic inflationary potential, still one the simplest inflation models, and consistent with observations when introducing dissipation in the system. The spectrum also depends on the statistical distribution function for inflaton fluctuations $n_*$ (whether in vacuum or thermal). We will focus on the vacuum case  $n_*=0$. Although the thermal case  can be also compatible with observations for the quartic chaotic model, typically it does not lead to enough amplification of the spectrum at the end. 

We want to maximise the amplification of the primordial spectrum between CMB scales and the end of inflation, and this may be achieved while being in the weak dissipative regime with $Q<1$ when observational constraints applied, but ending inflation in the SDR. Searching for this pattern motivates our choice of the dissipative coefficient and the inflationary potential. However, it must be stressed that WI all the way along in the SDR can provide a viable scenario to overcome the difficulties to have inflation (i.e. quasi de Sitter vacua) in string theory \cite{Motaharfar:2018zyb,das2018, das2019, arjunjaime,rudneibranden}, as given by the so-called swampland conjectures \cite{swampland0, swampland01, swampland1,swampland2,swampland3}. These demand the relative slope of scalar potentials to be larger than one in Planck units, i.e. having standard slow-roll parameters $\epsilon_\phi,\, |\eta_\phi| \gtrsim 1$, which invalidates slow-roll cold inflation. On the other hand, in WI those parameters are only required to be smaller than a factor $(1+Q)$, i.e., they can be larger than one in the SDR. Several examples of this can be found in the recent literature, depending on the combination of inflationary potential and particle interactions leading to dissipation: although a linear dissipative coefficient with a quadratic chaotic potential as in the WLI model is consistent with observations only for values $Q_* \lesssim 1$ \cite{WLI} at horizon-crossing, we can reach $Q_* \simeq O(100)$ with Higgs-like potentials \cite{WLImodels}; and a cubic dissipative coefficient as in MWI \cite{MWI} gives $Q_* \simeq O(100)$ when combined with a hybrid-like potential. In addition, the ``Transplanckian Censorship Conjecture'' \cite{TCC1, TCC2} sets an upper bound on the scale of inflation $V^{1/4} \simeq 3 \times 10^{-10} m_P$, in both cold and warm inflation \cite{dasTCC,arjunjaime}. This could be achieved with runaway potentials, but the SDR-WI version has the advantage of being already consistent with the other swampland conjectures \cite{rudneikamali, rudneidas}, and to solve the problem of the graceful exit \cite{rudneidas2}.

This works is organised as follows. We first review the basic of warm inflation, the dissipative coefficient and background evolution for the interaction pattern with light scalar dof \cite{SWLI} in Section II. In Section III we give the analytical expression of the growing mode for different constant values of $\kappa$. While this can be found in the literature for $\kappa=3,\,1,\,-1$ \cite{growing1,growing2,WLI,SWLI}, here we will generalize those results to other intermediate values. This will allow us to get the scalar primordial spectrum over the full range of inflation, and obtain the predictions for the scalar spectral index $n_s$ and the tensor-to-scalar ratio $r$. We comment also on the implications for PBHs at the end of Section III, and in Section IV we present the results for the spectrum of GW today. Finally, in section V we summarise and discuss our results. Details on the calculation and approximations of the dissipative coefficiente are given in Appendix A, while those related to the primordial spectrum are given in Appendix B. In Appendix C we provide a table with the parameter values used in this work, and the parameters for the fitting function of the primordial spectrum near the end of inflation. 

\section{Basics of Warm inflation: the ``Scalar'' WLI model}

After inflation ends, we must recover a radiation dominated universe before Big Bang Nucleosynthesis (BBN), made of minimum the Standard Model particles. This period when the inflaton energy density is transferred to the thermal bath is called reheating \cite{reheating1, reheating2, reheating}, and requires the interaction of the inflaton field with other particle species. But even before reheating, interactions can lead to dissipative effects during inflation, and the continuous transfer of inflaton energy density into radiation. In the context of slow-roll inflation, this is modelled by the introduction of an additional friction term, the  dissipative coefficient $\Upsilon$, in the inflaton $\phi$ equation of motion (eom):
\be
\ddot \phi + (3 H + \Upsilon) \dot \phi + V_\phi =0 \,,
\ee
where ``dot'' denotes derivative with respect to time, $H$ is the Hubble parameter, and $V_\phi= d V/d\phi$, $V$ being the inflationary potential. Energy lost by the inflaton is gained by the radiation energy density $\rho_r$:
\be
\dot \rho_r + 4 H \rho_r = \Upsilon \dot \phi^2  \label{rhor}\,.
\ee
where $\rho_r = \pi^2 g_* T^4/30$, $g_*$ being the effective no. of light dof. 
When friction dominates, either Hubble or dissipative, we enter in the slow-roll regime, and we can approximate the eoms by :
\bea
\dot \phi &\simeq& -\frac{V_\phi}{3 H(1 +Q)} \,, \label{phidotsl}\\
\rho_r &\simeq & \frac{3}{4}Q \dot \phi^2 \,, \label{rhorsl} 
\eea
where $Q=\Upsilon/(3H)$. The slow-roll conditions now read:
\bea
\epsilon_\phi &\simeq& \frac{m_P^2}{2}\left(\frac{V_\phi}{V}\right)^2 \ll 1 +Q \,,
\\
\eta_\phi &\simeq& m_P^2\frac{V_{\phi \phi}}{V}  \ll 1 +Q \,,
\eea
where $m_P$ is the reduced Planck mass. These conditions assume that $\rho_r \ll \rho_\phi=\dot \phi^2/2 + V  \simeq V$, which may be violated by the end of inflation. Indeed when $Q \gg 1 $ we will have $\rho_r \simeq \rho_\phi$ and inflation ends with a smooth transition to a radiation dominated universe. While studying the slow-roll regime we will work with the parameters $\eta_\phi$ and $\epsilon_\phi$, however to signal the end of inflation we will use instead the more accurate condition $\epsilon_H= - \dot H/H^2=1$,  with $3 H^2 m_P^2 = \rho_\phi + \rho_r$; the parameter $\epsilon_H$  tell us indeed whether the universe expansion is accelerated ($\epsilon_H < 1$) or not.

We will work with a quartic chaotic potential:
\be
V(\phi)= \frac{\lambda}{4} \phi^4\,,
\ee
and the dissipative coefficient given in a variant of the WLI  where the inflaton couples to a pair of scalars $\chi_{1,\,2}$ instead of fermions \cite{SWLI}. We can call this the ``Scalar Warm Little Inflation'' (SWLI) model. 
As in the original WLI, the inflaton is the relative phase between two scalars fields $\phi_{1,\,2}$ charged under a $U(1)$ broken gauge symmetry at a scale $M$, now coupled to a pair of scalars $\chi_{1,\,2}$ with coupling $g$. The system satisfies the interchange symmetry $\phi_1 \leftrightarrow i \phi_2$, $\chi_1 \leftrightarrow \chi_2$. This symmetry ensures that the scalar fields $\chi_1$, $\chi_2$ can be ``light'' during inflation, in the sense of having masses below the temperature bath, while avoiding large thermal radiative corrections to the inflaton potential. The interaction Lagrangian is given by:
\be
{\cal L}_I = \frac{1}{2} g^2 |\phi_1 + \phi_2|^2 |\chi_1|^2
   + \frac{1}{2} g^2 | \phi_1 - \phi_2 |^2 |\chi_2|^2 \,,
\ee
with $\phi_{1,\,2} = M e^{\pm i \phi/M}/\sqrt{2}$. Field dependent $\chi_i$ masses are then bounded by $gM$, but they may couple to other light fermion/scalar dof and acquire a thermal mass. The interchange symmetry requires both to couple to the same fields, and on average during inflation with $\phi \gg M$ we will have:
\be
m_\chi^2(T) \simeq \frac{g^2 M^2}{2} + \frac{h^2}{12} T^2+ \frac{h_S^2}{12} T^2\,,
\label{mchi2T}
\ee
where $h$ is their Yukawa coupling to fermions, and $h_S$ denotes a generic contribution from scalars to the thermal mass (including self-interactions). This extra contribution may help to keep the scalars in the high T-regime during inflation. Using standard tools from thermal field theory, the dissipative coefficient is given by (see Appendix \ref{Appdisscoeff}):
\bea
\Upsilon &=& \frac{4 g^2}{h^2} \cdot \frac{g^2 M^2}{T} F[m_\chi/T] \,, \label{Upsilon}\\
F[m_\chi/T]&=& \left(\frac{T}{m_\chi}\right)^3  \left( e^{-0.77 m_\chi/T} + 0.0135 h^6 e^{-20 T/m_\chi} \left( \frac{T}{m_\chi}\right)^5 \right) \,,
\eea
which holds when $T/H > 1$. In the high $T$ limit, $gM \ll T$, thermal corrections dominate the $\chi_i$ masses and  we have an inverse dissipative coefficient $\Upsilon \sim 1/T$; but as $T$ decreases during inflation, we may reach the low $T$ regime with a heavy scalar mediator, $m_\chi\sim gM \gg T$ and a dissipative coefficient $\Upsilon \sim T^\kappa$, with $\kappa \leq 7$. This will be relevant if we want to end inflation in the strong dissipative regime, $Q \gg 1$, where we may have second order induced GWs and eventually PBHs.

\begin{figure}[t]
  \centering
\begin{tabular}{ccc}
  \includegraphics[scale=0.35]{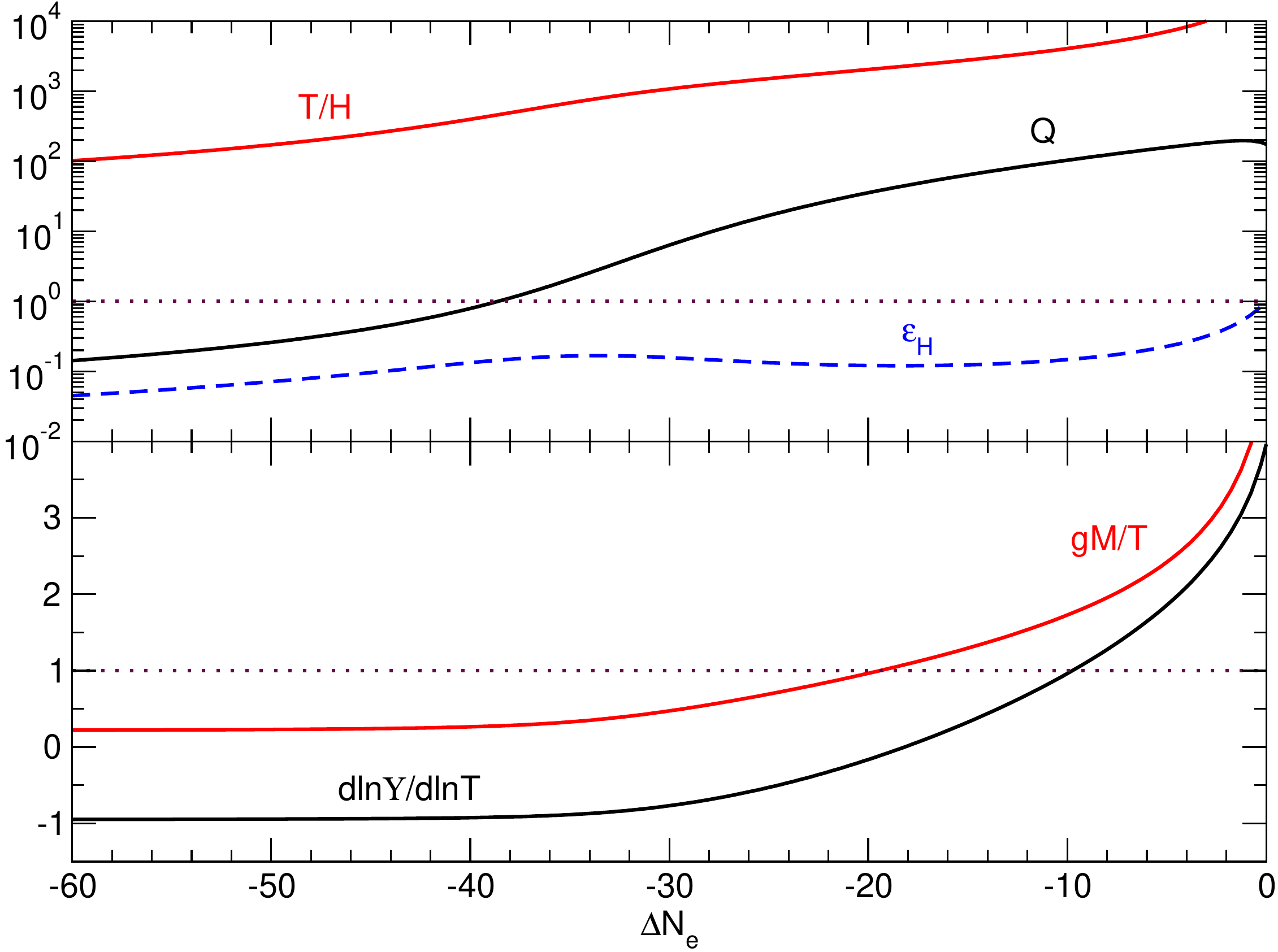} &&
  \includegraphics[scale=0.35]{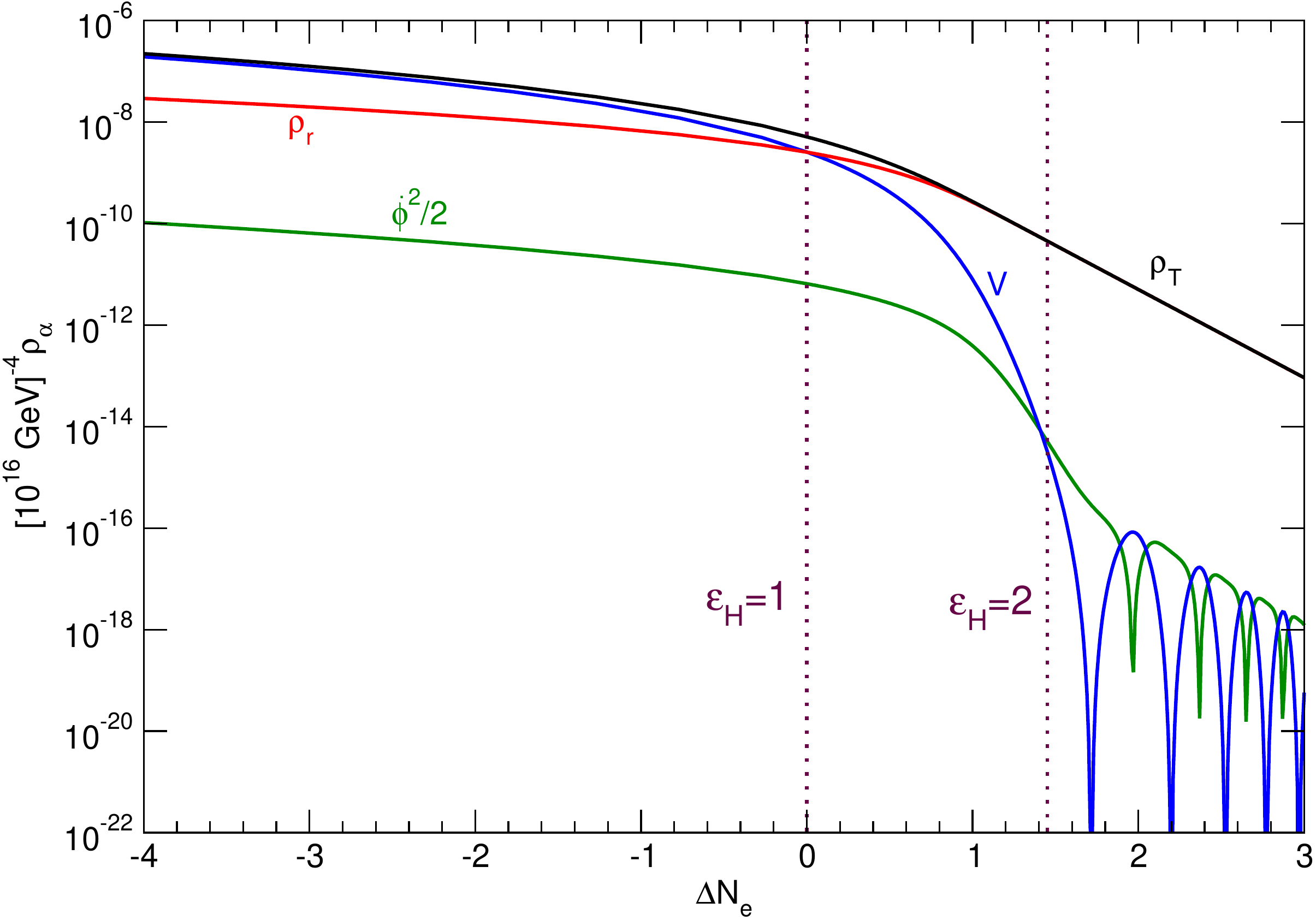}
  \end{tabular}
\caption{Left plot: evolution of the dissipative ratio $Q$, $T/H$ and $\epsilon_H$, and that of $gM/T$ and $d \ln \Upsilon / d \ln T = f_T -1$ (bottom panel), with respect to the no. of efolds left to the end of inflation.
Right plot: evolution of the energy densities (potential $V$ and kinetic $\dot \phi^2/2$ inflaton energy densities, radiation $\rho_r$ and total $\rho_T$) during the transition from inflation to a radiation dominated universe. The end of inflation happens at $\epsilon_H=1$, $\Delta N_e=0$, while it becomes RD at $\epsilon_H=2$. We have taken: $M= 10^{-4} m_P$, $g=1$, $h=2.5$, $h_S=4$, and $\lambda=10^{-14}$.} 
         \label{plotNedUpsgg1}
\end{figure}

In order to see when we may end in the strong dissipative regime, it is more useful to derive the slow-roll Eqs. for the dissipative ratio $Q$ and $T/H$, and the slow-roll parameter $\epsilon$. At first order in the slow-roll parameters we use the notation $\epsilon = \epsilon_H \simeq \epsilon_\phi/(1+Q)$.  Taking the derivative of $Q=\Upsilon/(3H)$ with respect to to no. of efolds $N_e$, using Eq. \eqref{Upsilon}, we have:
\be
\frac{d \ln T/H}{ dN_e} = \frac{1 }{1-f_T} \left(  ( 2-f_T) \epsilon - \frac{d \ln Q}{ dN_e}  \right) \label{THNe}\,,
\ee
where 
\be
f_T = \frac{d \ln F[m_\chi/T]}{d \ln T} \,. 
\ee
Combining now the derivative of the radiation slow-roll equation \eqref{rhorsl} with \eqref{THNe} we obtain:
\be
\frac{d \ln Q}{ dN_e} = \frac{2 (1+Q) }{5 - f_T + (3 + f_T)Q} \left(  (1 + f_T) \epsilon + (1-f_T) \eta \right) \label{QNe}\,,
\ee
with $\eta=\eta_\phi/(1+Q)$.  Finally, taking the derivative of $\epsilon \simeq \epsilon_\phi/(1+Q)$ together with Eqs. \eqref{phidotsl} and \eqref{QNe} we have:
\be
\frac{d \ln \epsilon}{ dN_e} = \frac{2}{5 - f_T + (3 + f_T)Q} \left( (2(5-f_T) + Q (5 + f_T)) \epsilon - (5- f_T + 4 Q) \eta \right) \,.
\ee
In particular for a quartic chaotic model one has $\eta_\phi = 3 \epsilon_\phi/2$, and
\bea
\frac{d \ln Q}{ dN_e} &=& \frac{(5-  f_T)(1+Q)}{5 - f_T + (3 + f_T)Q} \epsilon  \,, \label{dQN}\\
\frac{d \ln T/H}{ dN_e} &=& \frac{5-  f_T + (1+f_T)Q}{5 - f_T + (3 + f_T)Q} \epsilon  \,, \label{dTHN}\\
\frac{d \ln \epsilon}{ dN_e} &=& \frac{5 - f_T - 2 Q (1- f_T)}{5 - f_T + (3 + f_T)Q} \epsilon\,. \label{depsN}
\eea
The dissipative ratio always increases as far as\footnote{In order to have stable background evolution one requires $\Upsilon \propto T^\kappa$ with $\kappa \leq4$, $f_T \leq 5$, i.e., we may transfer energy to the thermal bath at a rate slower than the redshifting of standard radiation \cite{ianconsistency}. Once $\kappa > 4$ radiation will become dominant and inflation ends.} $f_T < 5$ , and so does $T/H$, but $\epsilon$ decreases when $Q > (5-f_T)/2(1- f_T)$, and again increases when $1 < f_T \leq 5$ for whatever value of $Q$. That means that to end inflation in the strong dissipative regime, $Q \gg 1$, we must be already in the low $T$ regime with $f_T > 1$ and $\Upsilon \propto T^\kappa$, $\kappa=f_T-1 >0$.
An example of  this behaviour is given in Fig. (\ref{plotNedUpsgg1}), where on the LHS we have plotted the evolution of the dissipative ratio $Q$, $T/H$, $\epsilon_H$, $gM/T$ and $d \ln \Upsilon / d \ln T = f_T -1$, with respect to the no. of efolds left to the end of inflation, taken as $\epsilon_H=1$. We have numerically integrated the background equations for the parameter values $M= 10^{-4} m_P$, $g=1$, $h=2.5$, $h_S=4$, $\lambda=10^{-14}$, and $g_*=12$. We recall that $M$ is $U(1)$ symmetry breaking scale, with the combination $gM$ given the non-thermal mass to the scalars $\chi_i$ coupled to the inflaton; the Yukawa couplings $h$ and $h_S$ on the other hand give their thermal mass (see Eq. \eqref{mchi2T}). The  dissipative coefficient Eq. \eqref{Upsilon} is proportional to $g^4M^2/h^2$, but only depends indirectly  on $h_S$ through the mass ratio $m_\chi/T$. Thus, hereon we will fix $h_S=4$ to ensure that thermal corrections to the scalar masses dominate say at least 50-60 efolds before the end of inflation, as shown on the LHS in Fig. (\ref{plotNedUpsgg1}). The slow-roll parameter $\epsilon_H \simeq \epsilon_\phi/(1 +Q)$ increases while the system is in the high $T$ ($g M/T < 1$ ) but weak dissipative regime ($Q < 1$). However when $Q \gtrsim 5/2$ around 35 e-folds before the end, the slow-parameter starts decreasing, but soon after we move into the low $T$ regime with $gM/T>1$, $\epsilon_H$ increases again and inflation ends.

Once inflation ends, dissipation quickly decreases, the inflaton starts oscillating around the minimum and on average its energy density behaves like radiation.
In addition, when ending inflation in the SDR the radiation energy density is already comparable to that of the inflaton field. The last few e-folds of inflation and the transition from inflation to radiation is shown on the RHS of Fig. (\ref{plotNedUpsgg1}), where the vertical dotted line labeled $\epsilon_H=2$ signals the time when the universe becomes radiation dominated, in about $O(1)$ efolds since the end. 

The behaviour of $\Upsilon(T)$, and in particular that of its derivative $d \ln \Upsilon/ d \ln T$, is relevant in order to get the expression for the primordial power spectrum: this controls how much the fluctuations in the radiation bath affect those of the inflaton field. In order to get the parameter space consistent with observations, i.e., the spectral index and tensor-to-scalar ratio, we need the expression for scalar primordial spectrum, which will be reviewed in the next section. Normalising the spectrum to the CMB value $P_{\cal R} = 2.1\times 10^{-9}$ at the pivot scale $k_*=0.05$ Mpc$^{-1}$, one fixes as usual one of the parameters of the model, i.e., the value of $\lambda \sim O(10^{-14})$.  In the example in Fig. (\ref{plotNedUpsgg1}) we have not used any information on the spectrum, and we have just taken $\lambda=10^{-14}$ as a typical value. The aim was to show the background evolution, which will not depend much on the particular value of $\lambda$. 

\section{Primordial spectrum in WI}

The primordial spectrum in WI is given by \cite{rudneisilva, chaoticwarm}:
\be
P_{{\cal R}_*} = \left( \frac{H_*^2}{2 \pi \dot \phi_*} \right)^2 \left( 1 + 2 n_* + \frac{T_*}{H_*}\cdot \frac{2 \sqrt{3} \pi Q_*}{\sqrt{3 + 4 \pi Q_*}}\right) G[Q_*] \,, \label{PRstar}
  \ee
where ``*'' denotes values at horizon crossing, and $n_*$ is the statistical distribution function of inflaton fluctuations; these may remain in vacuum ($n_*=0$), they may termalize due to the interactions ($n_* = (1 - e^{H_*/T_*})^{-1}$), or be in some other intermediate state. On the other hand, the function $G[Q_*]$ encodes what is called the ``growing mode'' \cite{growing1,growing2,WLI,SWLI}, due to the coupling of inflaton and radiation fluctuations. At linear order, it depends on the derivative $d \ln \Upsilon/ d \ln T$ at horizon crossing, such that when this is positive it does amplify the primordial spectrum, whereas it has the opposite effect for a negative derivative. When $d \ln \Upsilon/d \ln T=0$ the perturbation equations can be solved analytically at first order in the slow-roll parameters \cite{rudneisilva}, obtaining Eq. \eqref{PRstar} with $G[Q_*]=1$; otherwise the determination of $G[Q_*]$ requires the numerical integration of the system. This is important because $G[Q_*]$ and its derivative will directly enter in the expressions for the spectral index $n_s$ and the tensor-to-scalar ratio $r$.

  Tensor modes at linear order are not affected by dissipation, and their spectrum kept the standard form, with the tensor-to-scalar ratio given by:
  \be
  r= \frac{16 \epsilon_H}{(1+Q_*)}\cdot \frac{1}{F[T_*/H_*,Q_*]} \cdot \frac{1}{G[Q_*]} \,,  \label{tensortoscalar}
  \ee
   with
\be
F[T_*/H_*,Q_*] =1 + 2 n_* + \frac{T_*}{H_*}\cdot \frac{2 \sqrt{3} \pi Q_*}{\sqrt{3 + 4 \pi Q_*}} \,.
\ee
  The spectral index is given by the logarithmic variation of the primordial spectrum Eq. \eqref{PRstar} with the scale $k_*=a_* H_*$, which for superhorizon perturbations can be approximated by the derivative wrt the no. of efolds:
  \be
  n_s -1 = \frac{d \ln P_{{\cal R}_*}}{d \ln k_*} \simeq \frac{d \ln P_{{\cal R}_*}}{d N_e} = - 6 \epsilon_H + 2 \eta + (n_s-1)_{F} + (n_s-1)_{G} \,, \label{ns}
  \ee
where we have called:
\bea
(n_s-1)_{F}&=& \frac{d \ln F}{d N_e} = \frac{d \ln F}{d \ln Q_*} \cdot \frac{d \ln Q_*}{d N_e} + \frac{d \ln F}{d \ln T_*/H_*} \cdot \frac{d \ln T_*/H_*}{d N_e}\, \\
(n_s-1)_{G}&=& \frac{d \ln G}{d N_e} = \frac{d \ln G}{d \ln Q_*} \cdot \frac{d \ln Q_*}{d N_e}\,.
\eea

Our model depends on the parameters controlling the coupling of the inflaton to the scalars and their thermal masses: $M$, $g$, $h$ and $h_S$, and the self-coupling $\lambda$ for the inflaton. As usual, the normalization of the amplitude of the primordial spectrum with the Planck value fixes the value of $\lambda$, but for that we need the function $G[Q_*]$. We could scan over the parameters of the model, numerically integrate the perturbation equations, get the growing mode and the value of $\lambda$, and the predictions for the spectral index and tensor-to-scalar ratio. In practice, the function $G[Q_*]$ mainly depends on the value of $Q$ and $\kappa=d \ln \Upsilon/d \ln T$ at horizon crossing. Instead of scanning over the parameters $M,\,g,\,h,\,h_S$, we have run the perturbation equations for WI for a quartic chaotic potential and a generic dissipative coefficient $\Upsilon = C_\Upsilon T^\kappa$, for different values of constant $\kappa$; varying $C_\Upsilon$ we tune the value of  $Q_*$. In the numerical simulations we have set $\lambda=10^{-14}$ and read the background values $Q_*$, $H_*$, $T_*$, etc...60 e-folds before the end of inflation. In order to get the growing mode $G[Q_*]$, we compare the amplitude of the spectrum at the end of inflation with Eq. \eqref{PRstar} with $G[Q_*]=1$ (see Appendix \ref{growing}). Depending on the sign of $\kappa$, the numerical results can be well fitted by the functions:
\bea
G[Q,\kappa] &=& (1 +  e^{\alpha_s} Q^{\beta_s} + e^{\alpha_w} Q^{\beta_w})^\kappa \,, \;\;\; \kappa > 0 \,, \label{GQpos}\\
G[Q, \kappa] &=& \frac{(1 + a_0 Q^{a_1})^{a_5}}{(1 + a_2 Q^{a_3})^{a_4}} \,, \;\;\; \kappa \leq 0 \label{GQneg}\,,
\eea
where the coefficients are also functions of $\kappa$ and they are given in Appendix \ref{growing}. 

Finally, in order to get the predictions for $n_s$ and $r$ and compare with Planck data, we need to get the no. of efolds at which the Planck pivot scale $k_*=0.05$ Mpc$^{-1}$ leaves the horizon during inflation. This is done using the standard relation:
\be
\frac{k}{a_0 H_0} = \frac{a_k H_k}{a_{end} H_{end}} \cdot \frac{a_{end} H_{end}}{a_{RH} H_{RH}}\cdot \frac{a_{RH} H_{RH}}{a_{eq} H_{eq}}\cdot \frac{a_{eq} H_{eq}}{a_0 H_0} \,,
\ee
where ``0'' denotes present values and ``{\it eq}'' denotes the time of matter-radiation equality; ``{\it RH}'' signals the end of reheating,  and ``end'' means the end of inflation, i.e. when $\epsilon_H=1$. For any comoving scale $k$, one obtains for the no. of efolds at horizon crossing \cite{Noefolds1, Noefolds2, Noefolds}:
\be
N(k)= 56.01 - \ln \frac{k}{k_P} + \ln \frac{a_{end}}{a_{RH}} + \ln \frac{\rho_{RH}^{1/4}}{10^{16} \gev} + \frac{1}{2}\ln \frac{\rho_k}{\rho_{RH}} \,. \label{Nefolds}
\ee
This value depends on the details of reheating, and how the universe becomes radiation dominated such that its expansion is dominated by a fluid with equation of state $w=1/3$. However in our scenario when we en inflation in the SDR (our case of interest) radiation in the sense of a thermal bath quickly dominates as shown on the RHS in Fig. (\ref{plotNedUpsgg1}). Therefore, without further assumptions about the interactions of the inflaton field and the need of introducing any other decay channel during reheating, we only need approximately a couple of e-folds after the end of inflation to recover our RD universe, which happens when $\epsilon_H=2$. The total energy density at this point is the value $rho_{RH}$ in Eq. \eqref{Nefolds}.


\begin{figure}[t]
  \centering
\begin{tabular}{ccc}
  \includegraphics[scale=0.35]{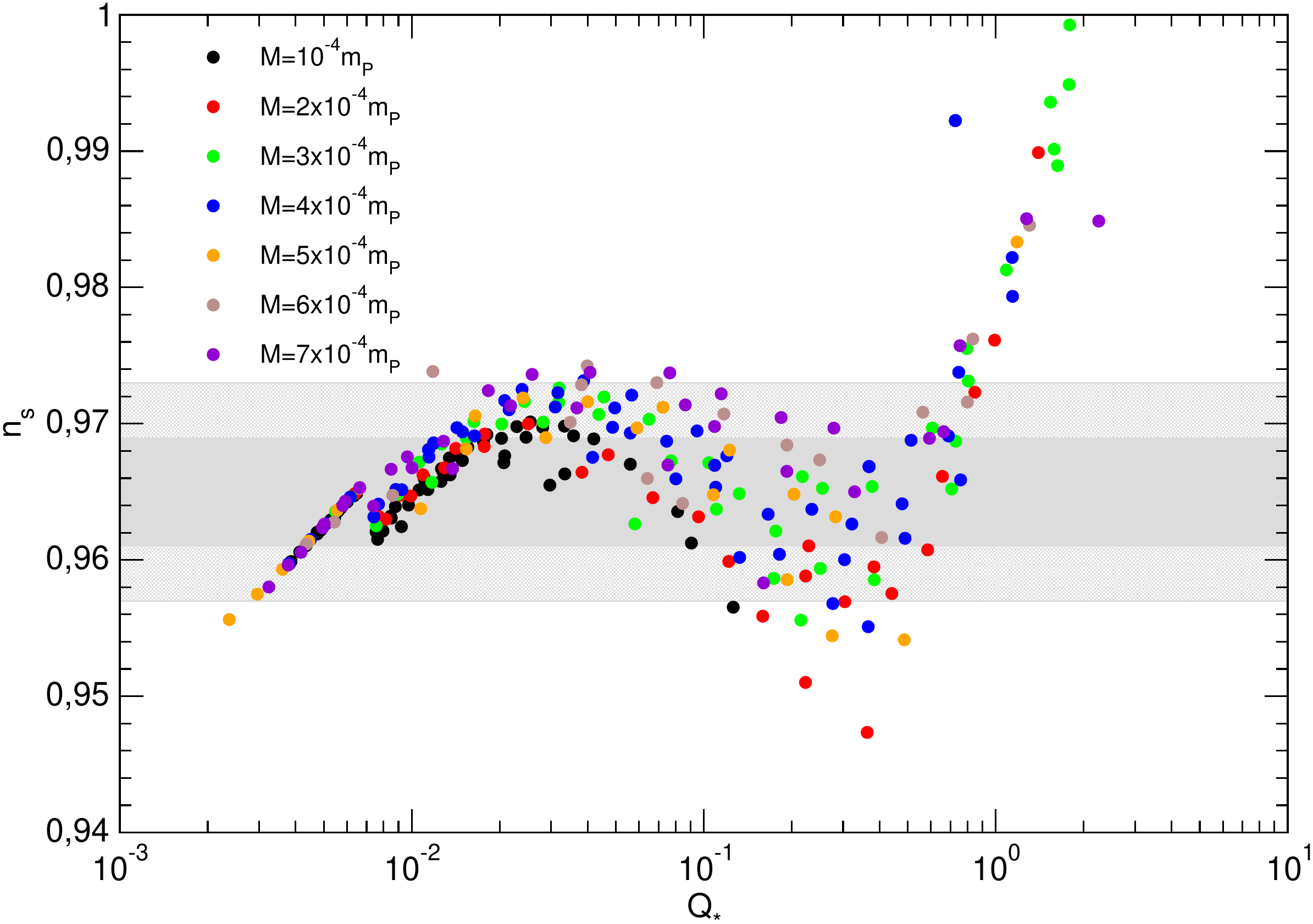} &&
  \includegraphics[scale=0.35]{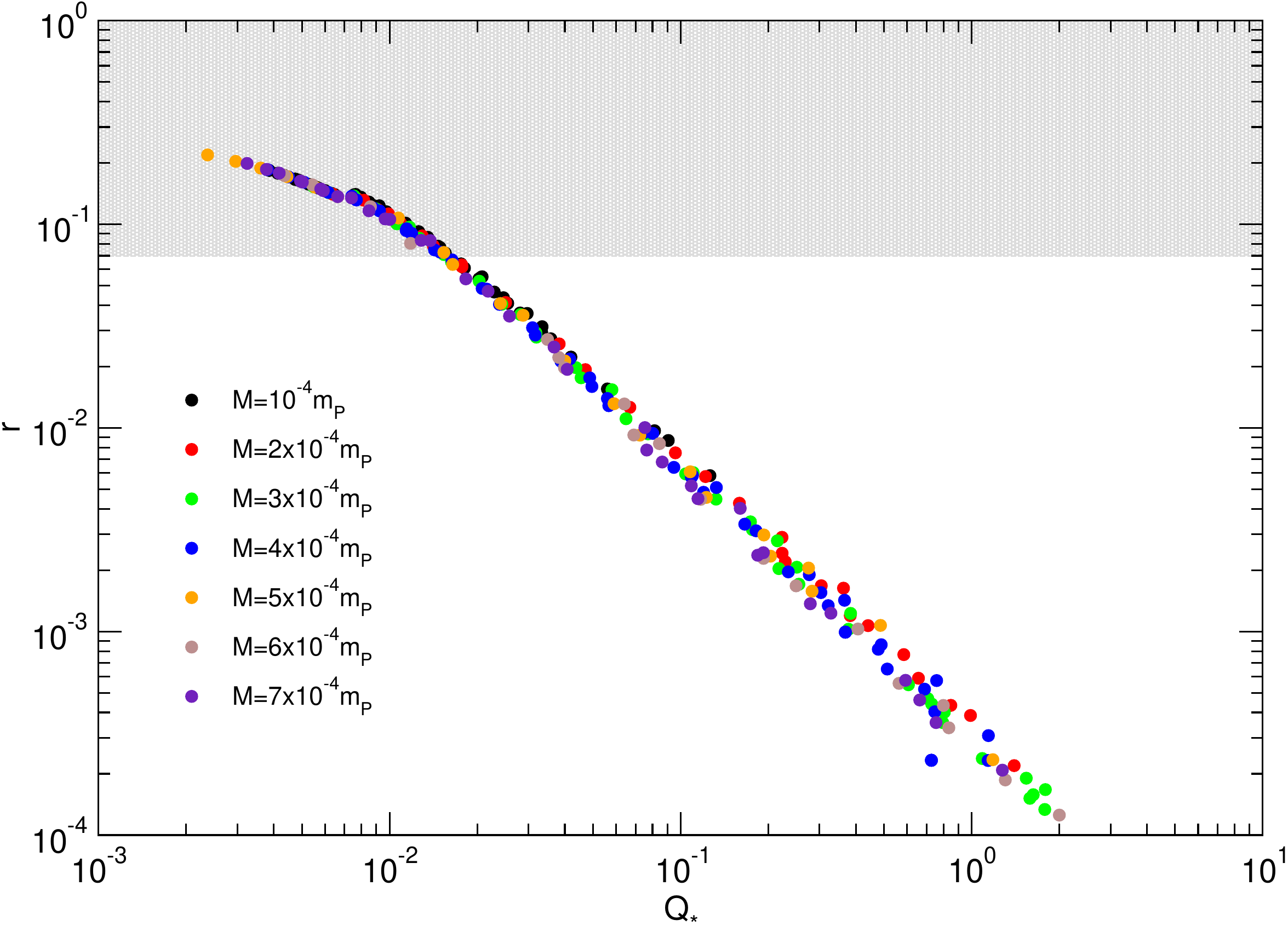} 
  \end{tabular}
  \caption{Left: prediction for the spectral index versus the value of the dissipative ratio $Q_*$; the grey shaded regions indicated the 1$\sigma$ (dark) and 2$\sigma$ (light) Planck limits. Right: tensor-to-scalar ratio for the same parameter values; the grey shaded region is excluded by observations, $r < 0.07$. As parameter values, we have taken the inflaton self coupling  $\lambda=10^{-14}$, the symmetry breaking scale $M$ as indicated in the plot, vary the coupling of the scalars to the inflaton $g$ and fermions $h$, and fixed their coupling to other species $h_S=4$.} 
         \label{plotnsr}
\end{figure}

We can now scan over the model parameters, $M$, $g$, $h$ and $h_S$ to get the predictions. From the background evolution, we get the values of $H_*$, $Q_*$, $T_*/H_*$ and $\kappa_*$, at $N(k_*)$ given in Eq. \eqref{Nefolds}; normalising the spectrum Eq. \eqref{PRstar} with $n_*=0$ to the Planck value $P_{\cal R} = 2.1\times 10^{-9}$ at the pivot scale we fix the value of $\lambda$; and from Eqs. \eqref{ns} and \eqref{tensortoscalar} the predictions for the model. Those are shown in Fig. \eqref{plotnsr}, where we have scanned over different values of $g$ and $h$ in the range [0.1,4], for the values of $M$ indicated in the plot, and taking $h_S=4$ as an example.  The spectral index is consistent with observations for a dissipative ratio $ 3\times 10^{-3} \lesssim Q_* \lesssim 1$, but the tensor-to-scalar ratio requires $Q_* \gtrsim 10^{-2}$. As usual in WI, the larger the value of $Q_*$, the more suppressed is $r$. But we cannot go beyond $Q_* \gtrsim 1$ because the spectral index becomes too large (and eventually blue-tilted) in this model.  

\begin{figure}[t]
  \centering
\begin{tabular}{ccc}
  \includegraphics[scale=0.35]{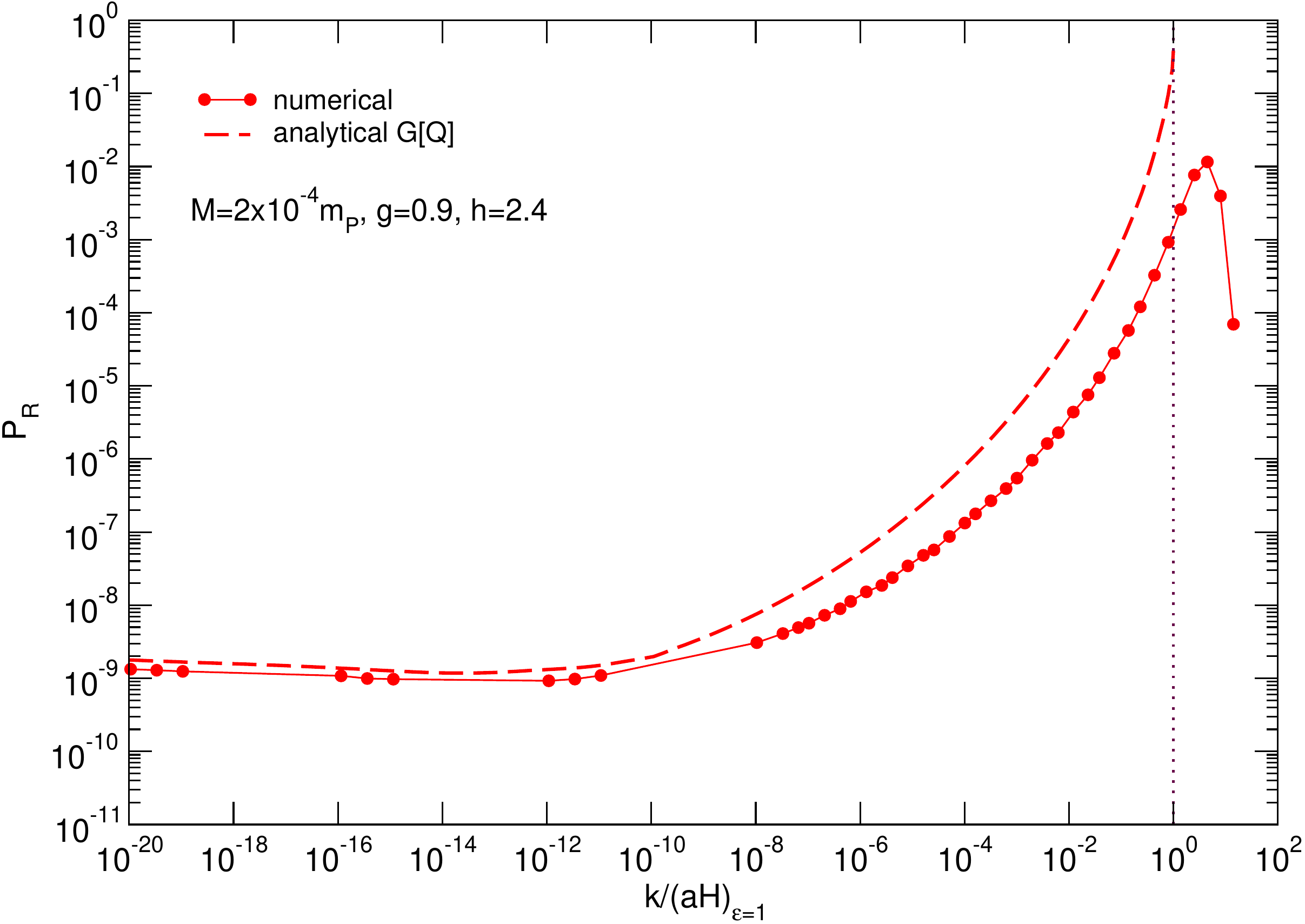}  &&
  \includegraphics[scale=0.35]{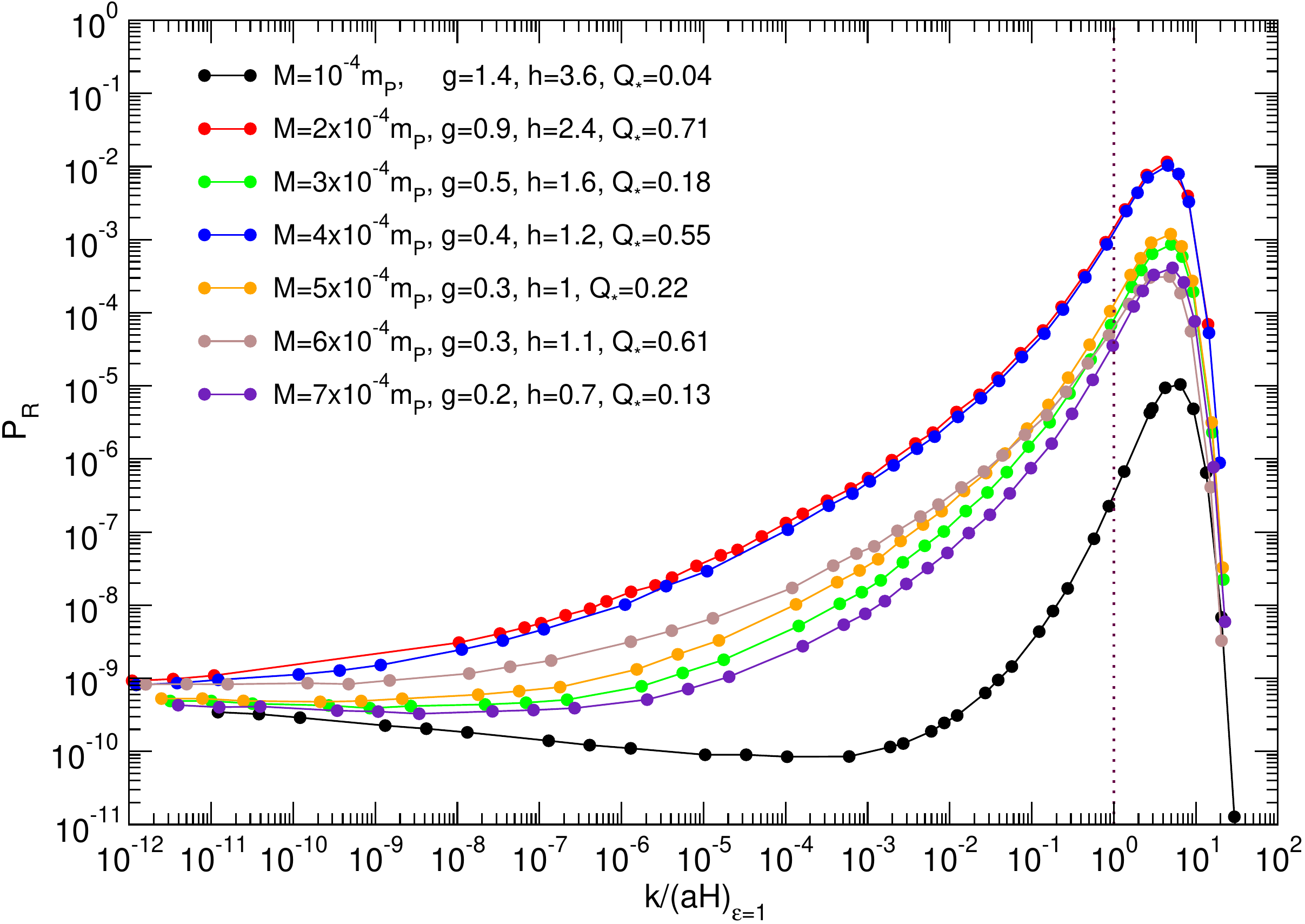} 
\end{tabular}
\caption{Left plot: comparison of the semianalytical primordial power spectrum $P_{\cal R}$ (dashed-lines) with the numerical results (filled circles), with respect to the comoving wave-number normalised by the value of $(aH)$ at the end of inflation, for the parameter values indicated in the plot. Right plot: $P_{\cal R}$ spectrum obtained numerically for the set of parameters given in the plot.
$M$ is the symmetry breaking scale, $g$ the inflaton coupling to the scalars $\chi$, $h$ the Yukawa coupling of the scalars to fermions. The inflaton self-coupling $\lambda$ in each case is adjusted imposing the normalization of the primordial spectrum to the Planck value at the pivot scale. The value of the dissipative ratio $Q_*$ when the pivot scale crossed the horizon is also included.  
}
  \label{plotPR}
\end{figure}

This sets the parameter space compatible with observations, and the condition $Q_* \gtrsim 10^{-2}$ ensures that we end inflation in the SDR, with $\kappa > 0$ and a spectrum which amplitude is more amplified towards the end due to the effect of the thermal fluctuations. However, while our seminalatycal estimation of the growing mode function $G[Q_*]$ works well in this model when $\kappa \lesssim 0$, starts failing toward the end of inflation and tends to overstimate the amplitude of the spectrum. This can be seen on the LHS plot in Fig. \eqref{plotPR}, where for a set of parameters ($M$, $g$, $h$) we compare the amplitude of the primordial spectrum $P_{\cal R}[k]$ obtained analytically with the approximation for $G[Q]$ (dashed-lines) with the result of numerically integrating the equations for the perturbations (filled circles). The value of the wavenumber $k$ on the X-axes is normalised by $aH$ at the end of inflation when $\epsilon_H=1$, i.e., the last mode that exits the horizon. The seminalytical approximation is only applicable to those modes, and because of that the dashed line end at $k_{\rm end}=(aH)_{\epsilon_1}$. On the other hand, the numerical integration allows us to read the spectrum at the end of inflation for all modes. And indeed, because during WI modes are amplified before they cross the horizon, the maximum of the spectrum is obtained for slightly larger modes with $k \simeq 5 k_{\rm end}$. Larger modes will not have time to be amplified, and the spectrum falls practically exponentially afterwards. We think that the failure of our approximation for the growing mode function near the end of inflation is due to the fact that the dissipative coefficient $\Upsilon$ changes its behavior but with a faster varying  $\kappa=d  \ln \Upsilon/d \ln T$ (see Fig. \eqref{plotNedUpsgg1}). But in the regime when $\kappa \simeq $ Constant our estimations works well. In particular this is the case  for modes that exit the horizon $O(60)$ efolds before the end, i.e $k\simeq 10^{-19} k_{\rm end}$ in Fig. \eqref{plotPR}, and we can rely on that to get the values of the spectral index and tensor-to-scalar ratio. Although our function $G[Q_*]$ overestimates the amplitude of the spectrum at the end, we can use it as an indication of which parameter values can lead to the amplification of the spectrum. We have then selected some set of parameters consistent with the observations of the spectral index and tensor-to-scalar ratio, and numerically integrate the Eqs. for background and perturbations. Typically, to be consistent with observations we need (a) $M \simeq O(10^{-4})m_P$, (b) to  keep $h_S=4$ in order to ensure the high $T$ regime, i.e., $m_\chi/T \ll 1$ when observational constraints on the primordial spectrum applied, and (c) to choose different values of the couplings $g$ and $h$ in order to have $Q_* \sim O(10^{-1}-1)$ but ending inflation in the SDR. The amplitude of the spectrum at the end of inflation is shown on the LHS in Fig. \eqref{plotPR}. The smaller amplitude at the end correspond to the example with the smaller $Q_* \simeq 0.04$, which enters later in the SDR. 

\begin{figure}[t]
  \centering
  \includegraphics[scale=0.35]{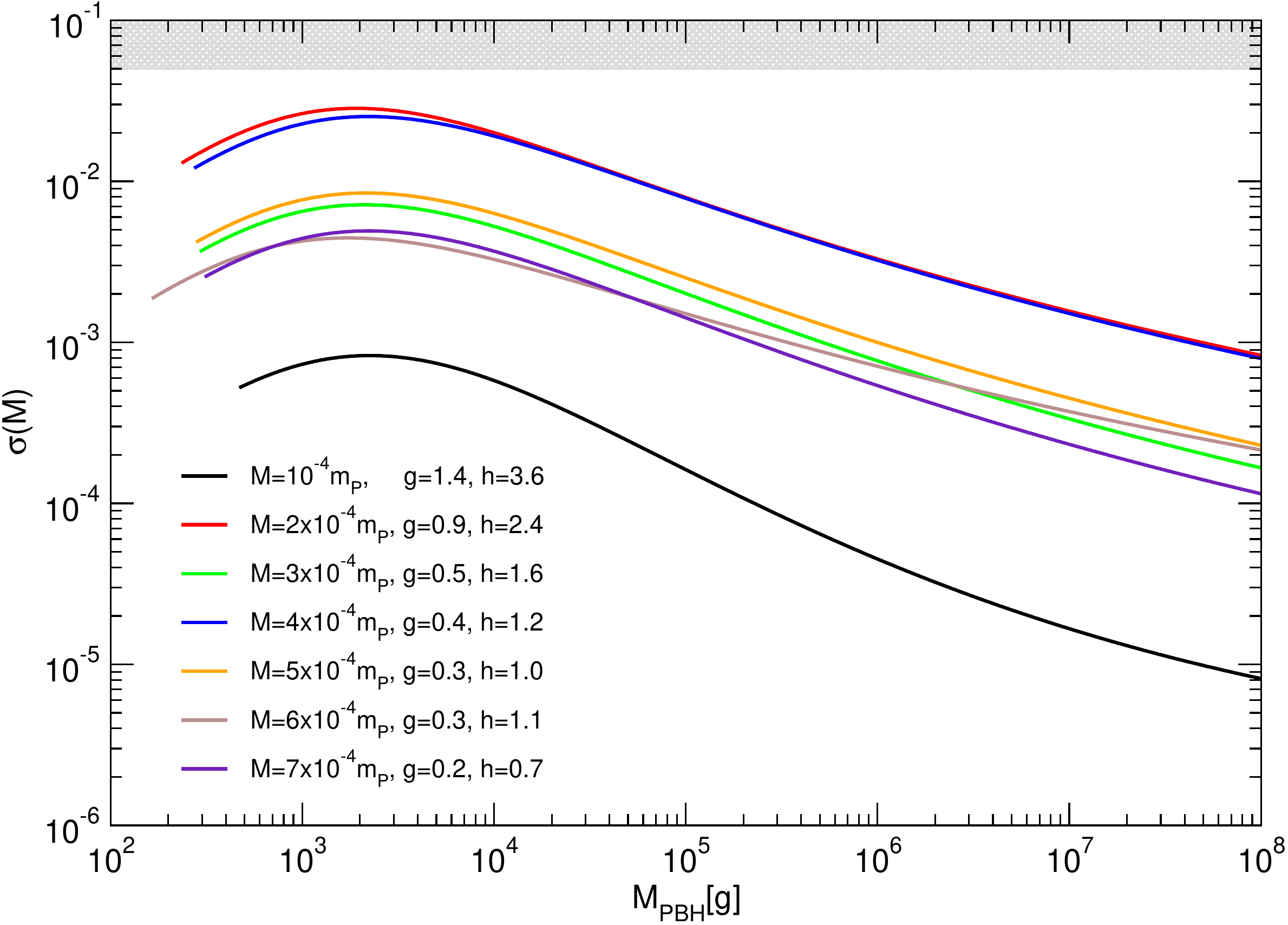} 
  \caption{ Variance of the density fluctuations with respect to the PBH mass, for the same parameter values than in Fig. \eqref{plotPR}. The shaded area shows the region $\sigma(M) > 0.05$. } 
  \label{plotsigmaM}
\end{figure}

  In all the cases, the spectrum is amplified before the end, when $\kappa$ turns positive. We reach values $P_{\cal R} \approx O(10^{-2}-10^{-4})$ that may lead to PBH formation  due to the collapse of the over-dense perturbations, with a mass given as a fraction $\gamma \simeq 0.2$ of the horizon mass at re-entry \cite{PBH1, PBHGreen}:   
\bea
M_{PBH} (k) &=& \gamma \frac{4 \pi}{3} \frac{\rho}{H_M^3} = \gamma \frac{4 \pi m_P^2}{H_M} \nonumber \,, 
\eea
where $H_M$ is the Hubble parameter at the time of formation. Because we only get a large amplitude of the spectrum by the end of inflation, PBHs may will form during the RD era that follows inflation, with \cite{PBHDrees}:
\be
M_{PBH}[\rm g]= \frac{\gamma}{(5.4185\times 10^{-24})^2} \left(\frac{3.36}{g_*}\right)^{1/3} \left(\frac{k_{\rm end}}{k_M}\right)^2 \frac{1}{k_{\rm end}^2} \,. \label{Mkend}
\ee
where $g_*\simeq 106.75$ is the effective no. of dof at re-entry, $k_M=a_M H_M$ and $k_{\rm end} \simeq O(10^{20})$ Mpc$^{-1}$ in our models. This gives PBH masses $M_{PBH} \lesssim 10^6 ~{\rm g}$, that will therefore evaporate before BBN \cite{PBHCarr, PBHCarr2, PBHCarr3, PBHCarr4}. Still, they can produce relics that might overclose the universe, and if one consider this (model dependent) possibility, it may impose an upper limit on their mass fraction $\beta(M) \lesssim 10^{-16}$. Using the Press-Schechter formalism, for a Gaussian distribution of fluctuations $\delta=\delta \rho/ \rho$ the mass fraction is given by:
\be
\beta(M) = \int_{\delta_c}^\infty d\delta P(\delta) = \frac{2}{\sqrt{2 \pi} \sigma(M)} \int_{\delta_c}^\infty d\delta {\rm exp}( -\frac{\delta^2}{\sigma(M)^2}) = {\rm erfc}(\frac{\delta_c}{\sqrt{2}\sigma(M)})  \,, \label{betaM}
\ee
where ``erfc'' is the complementary error function, $\delta_c$ is the critical density contrast, and $\sigma(M)$ the variance of the density fluctuations at a mass scale $M$. When PBH formation takes place in a RD universe, $\delta_c \simeq 0.414$ \cite{Harada}. The variance of the density fluctuations at a mass scale $M$ is given in terms of the primordial spectrum as:
\be
\sigma^2(M)= \frac{4 (1+w)^2}{(5+3w)^2} \int d\ln k ~\left(\frac{k}{k_{M}}\right)^4 W^2[k/k_M] P_{\cal R}(k) \label{sigmaM}\,,
\ee
where $W[k/k_M]$ is a Gaussian window function that smooths the perturbation on the comoving scale $k_M$. In order to get an approximation of how large can be the primordial spectrum by the end of inflation in order not to be in conflict with limits on evaporating PBHs, we could parametrize the primordial spectrum around the scale of interest as a power-law, $P_{\cal R} (k) \simeq P_{\cal R}(k_M) (k/k_M)^{(n_s-1)}$. The function $\sigma(M)$ is then given by \cite{PBHDrees}:
\be
\sigma(M) \simeq \frac{4}{9\sqrt{2}} C(n_s) \sqrt{P_{\cal R}(k_M)} \,,
\label{sigmaMapprox}
\ee
where $C(n_s)=\Gamma^{1/2}[(n_s+3)/2]$, and it ranges between $1 \lesssim C(n_s) \lesssim 1.4$ for $1 \leq n_s \leq 3$. The limit $\beta(M) \lesssim 10^{-16}$ translates into $\sigma(M) \lesssim 0.05$, and using the approximation in Eq. \eqref{betaM}, this gives $P_{\cal R} \lesssim 0.02$. However, our spectra are not well approximated by a simple power-law with constant spectral index, with the running and the running of the running being important during the last 10-20 efolds, and the approximation in \eqref{sigmaMapprox} tends to overestimate $\sigma(M)$. For the numerical spectra in Fig. \eqref{plotPR} we found that for $k \geq 10^{-12} k_{\rm end}$ they can be parametrised as:
\be
\ln P_{\cal R} [k] = c_0 + (1 - {\rm exp}(b_0 x -b_1)) P_5[x] \,,
\ee
where $x=k/k_{\rm end}$, $P_5[x] = \sum_{i=0,5} a_i x^i$, and we have chosen the common reference value $c_0=-25$ for all the spectra for convenience. Values of the different parameters $b_i$, $a_i$ are given in Table \ref{tableparamaibi} in Appendix \ref{param}. With this semianalytical approximation for the spectrum, we have obtained the variance $\sigma(M)$ by numerically integrating Eq. \eqref{sigmaM}; the results are plotted in Fig. \eqref{plotsigmaM}, for our selected values of parameters. In all the cases we are below the potential limit on $\sigma$, with a negligible mass fraction $\beta(M) \ll 10^{-16}$.

\section{Second order induced spectrum of Gravitational Waves}

Although the enhancement of the primordial scalar spectrum only leads to light, evaporating PBHs, it will act as an efficient second order source for the primordial tensors \cite{tomita, Matarrese:1992rp,GWMollerach, GWWands, GWBaumann, GWEspinosa,GWKohri,GWInomata}. To compute the power spectrum of the GW energy density today we follow the semi-analytical approach given in Ref. \cite{GWKohri}. The energy density parameter for GW per logarithmic $k$ interval today is given by:
\be
\Omega_{GW,\,0} = \frac{\rho_{GW} (k)}{\rho_{T,\,0}} = 0.39 \left( \frac{g_* (T_c)}{106.75} \right)^{-1/3} \Omega_{r,\,0} \Omega_{GW}(k, \tau_c)  \label{OmegaGW0}\,,
\ee
where $\Omega_{r,\,0} h^2 \simeq 4.18 \times 10^{-5}$ is the radiation density parameter today, $\tau$ the conformal time, the subindex ``$c$'' denotes the time when the perturbation is well inside the horizon after re-entry during RD, and $g_*(T_c)$ is the effective no. of relativistic dof at that time. The GW spectral density is given in terms of the tensor spectrum:
\be
\Omega_{GW} (k, \tau) = \frac{1}{24} \left( \frac{k}{a H} \right)^2 \overline{P_h ( k, \tau)}  \label{OmegaGW}\,,
\ee
and the induced tensors at second order\footnote{First order metric tensor perturbations decouple from scalar and vector perturbations, and are gauge independent. That is not the case for second order induced metric tensor perturbations, which are generically gauge dependent \cite{Hwang:2017oxa,DeLuca:2019ufz,Inomata:2019yww}. However, it has been shown that several choices of gauge, like the synchronous gauge, the uniform curvature gauge,  and the Newton or zero-shear gauge, yield the same GW spectrum today induced by scalar perturbations during radiation \cite{DeLuca:2019ufz,Inomata:2019yww, Domenech:2020xin}.} by the primordial scalar fluctuations by:
\be
\overline{P_h (k ,\tau)} = 4 \left(\frac{4}{9} \right)^2 \int_0^\infty dv \int_{|1-v|}^{1+v} du
\left( \frac{4 v^2 - (1 + v^2 -u^2)^2}{4 uv}\right)^2 P_{\cal R} (k v) P_{\cal R} (k u) \overline{I^2_{RD}(u,v,k\tau)} \label{Ph}\,,
\ee
where the analytical expression for the time average $\overline{I_{RD}^2}$ can be found in \cite{GWKohri}.

\begin{figure}[t]
  \centering
  \includegraphics[scale=0.4]{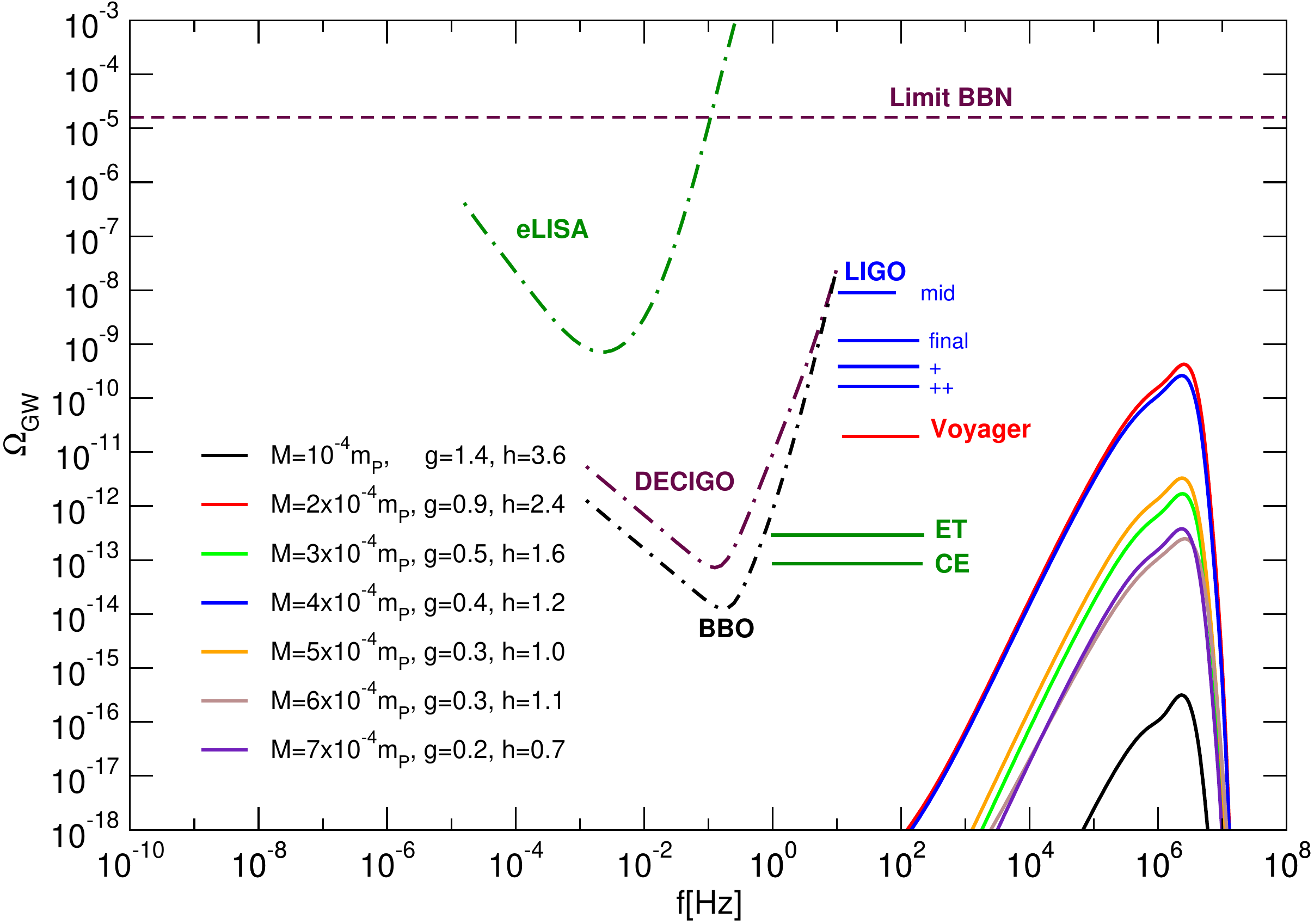} 
  \caption{Spectral density of GW today for the SLWI model with a quartic chaotic potential, for parameter values as indicated in the plot. The sensitivity curves for some GW detectors are also included. The horizontal dashed line is the BBN limit.} 
         \label{plotGW}
\end{figure}

Finally, in Fig. (\ref{plotGW}) we have the spectral density of induced GW today for our model and the set of parameters considered in this work, obtained from Eqs. \eqref{OmegaGW0}-\eqref{Ph}. The sensitivity curves for present and planned GW detectors \cite{Mandic, decigo, CE, ET, eLisa} are included as a reference. Also the BBN limit \cite{GWKohri}, $\Omega_{GW} h^2< 1.8 \times 10^{-6}$,  with $h=0.6736$, that sets the amount of extra radiation that can be present at the time of formation of light nuclei. The spectrum is amplified due to (a) modes that exit the horizon near the end of inflation, re-entering soon after during RD; (b) modes upto $k \sim 5 k_{\rm end}$ that never become super-horizon, but gives the maximum amplification at the end of inflation. Those modes are already inside the horizon when the RD era starts. Typically we have $k_{\rm end} \simeq O(10^{20})$ Mpc$^{-1}$, and this tranlates  into the frequency range today $f\simeq O(10^5-10^{6})$ Hz for the peak of the spectrum, with $\Omega_{GW,\,0} \approx 10^{-9}$ for the maximal amplification with the parameter values considered. The spectrum falls exponentially after the maximum, as it does the primordial spectrum that sources it, but it has a slope slightly smaller than the standard $f^3$ before the maximum. Indeed, the reduction in the slope is due to log corrections and the spectrum behaves as $\Omega_{GW}(f < f^{\rm peak}) \propto f^{3}\ln^2 f/f^{\rm peak}$, with $f^{\rm peak} \sim 2\times 10^6$ Hz \cite{Yuan:2019wwo,Domenech:2021ztg}.
Although the values of the amplitude and frequency peak leave these waves so far outside the range of detection of current and near future GW experiments, far future GW experiments sensitive to frequencies larger than kHz might be able to detect them. This would require develolping new GW detectors like those proposed for example in  \cite{Aggarwal:2020umq, Herman:2020wao}.  

For the model parameters considered, the primordial scalar spectrum is not larger than $O(10^{-2})$. It would be interesting to find parameter values for which the spectrum is closed to the perturbative limit at the end $P_{\cal R} \simeq O(0.1-1)$ and obtained the maximum possible amplification. Given that GW spectrum scales with $P_{\cal R}^2$, this would enhance the GW spectrum roughly 3-4 orders of magnitude with respect to the maximal value found in Fig. \eqref{plotGW}, close to the BBN limit. Although the frequency range of the maximum will still be far from present and near future GW detectors, smaller frequencies may enter within the reach of the ``Einstein Telescope'' or the ``Cosmic Explorer'' \cite{CE,ET} (labelled ET and CE respectively in Fig. \eqref{plotGW}). But in order to efficiently identify the parameter values $M$, $g$ and $h$ for the maximal amplification, further work is first needed to  characterize the enhancement of the spectrum at the end of inflation with a varying vaue of $\kappa$, which we defer to the future.

Similarly, the peak frequency depends on the value of $k_{\rm end}$, i.e, on the value of $H_{\rm end}$ and the energy scale at which inflation ends. For the quartic model studied here this is of the order $H_{\rm end} \simeq O(10^{9})$ GeV. Models with a smaller inflationary scale, like hybrid models or hilltop-like models might bring the spectrum within the detectable frequency range. However for this kind of models the relevant 50-60 efolds of inflation typically take place already in the SDR, and the relative amplification of the spectrum towards the ends diminish. We think that chaotic models are best suited to explore the kind of large amplification of primordial perturbations needed to impact on the GW spectrum, but this is of course a model dependent question and other possibilities are not excluded.


\section{Conclusions}

The main result of this work is given in Fig. \eqref{plotGW}, the spectral density of GW today for a model of quartic chaotic warm inflation. Those are induced at second order by the enhancement of the primordial spectrum of curvature perturbations. In warm inflation we have a coupled, resonant system of inflaton and radiation perturbations, and when dissipation $\Upsilon$ grows with the $T$ of the thermal bath, this gives rise to the amplification of the primordial perturbations \cite{growing1, growing2}. When this effect already takes place at large scales scales around the pivot value at which the primordial spectrum is normalised, it may lead to a blue-tilted primordial spectrum excluded by observations. To avoid this, we have focused on the SLWI model \cite{SWLI}, a variant of the LWI model with an inverse $T$ dependent dissipative coefficient at CMB scales, due to the coupling of the inflaton to a pair of light scalars. However, as inflation proceeds, scalar masses become larger than $T$, we move into the low-$T$ regime for dissipation, and we recover a model with $\Upsilon$ growing with $T$. The spectrum is then naturally enhanced by the end of inflation. We combine this with a quartic chaotic potential, instead of the quadratic one studied in \cite{SWLI}: while a quadratic potential is compatible with observations when the system is already in the SDR $50-60$ efolds before the end, the amplification experience by the spectrum upto the end of inflation is larger in the quartic model. 

By numerically integrating the perturbations Eqs. in a toy model with a quartic potential and $\Upsilon \propto T^\kappa$, with $\kappa=$Constant, we have obtained seminalytical expressions for the amplitude of the primordial spectrum including the so-called growing mode function, $G[Q_*]$, for different values of $\kappa$. While this works well for the model considered in this work at O(50-60) efolds before the end of inflation, when the variation of $\kappa$ can be neglected, the approximation fails towards the end and tends to overestimate the spectrum. Nevertheless, with the results at O(50-60) efolds we scanned over the parameters of the model $M$, $g$ and $h$, and obtain the parameter space compatible with (a) the observational limits on the spectral index and tensor-to-scalar ratio, (b) ending inflation in the SDR. Beyond that and near the end of inflation, we would need to characterize and understand better our spectrum, including the effects of the variation\footnote{Integrating the perturbations with constant values of $\kappa$ all along inflation we have checked that our fitting functions holds even for modes exiting the horizon a few efolds before the end.} of $\kappa$. This is beyond the scope of this work, where we try to estimate first whether the mechanism for the second order enhancement of GW works. Then we have selected some model parameters that we think can be representative of the mechanism, to compute numerically the spectrum of scalar perturbations and the induced GW spectrum. 

This amplification may also give rise to the formation of PBHs when the perturbations re-enter the horizon, soon after the end of inflation. Nevertheless, we only have light, evaporating PBHs with masses $M_{PBH} \lesssim 10^6$ g. They may give rise to Planck relics that may overclose the universe. We have compute the variance function $\sigma(M)$, which for the cases considered is always small. This translates into a negligible mass fraction and no further constraint on this set of model parameters. 

The maximal amplitude of GWs is obtained at frequencies $f \simeq 0.1-1$ MHz, outside the range of detection of GW detectors, but close for example to the typical stochastic spectrum obtained during preheating after inflation \cite{GWpreh1, GWpreh2, GWpreh3, GWpreh4, GWpreh5}. We notice that we do not expect any additional contribution due to preheating, i.e., we do not expect any parametric resonance or additional particle production to take place at the end of inflation. For the parameter space that amplify the spectrum in our case, inflation ends in a smooth transition from inflation to a radiation dominated universe, with a damped, subdominant oscillating inflaton. In any case, our mechanism adds to the possible cosmological sources of a background of gravitational sources beyond MHz frequencies, frequencies for which no astrophysical source has been identified, and which may be searched for in future detectors \cite{Aggarwal:2020umq, Herman:2020wao,GWMHz}.

\appendix

\section{Dissipative coefficient in the SWLI}
\label{Appdisscoeff}
Applying standard tools in Thermal Quantum Field Theory, the dissipative coefficient due to the coupling of the inflaton field to a pair of scalars $\chi_i$ is given by \cite{mossdiss,ramosdiss,joaodiss}:
\be
\Upsilon = \frac{4 g^4 M^2}{T} \int \frac{d^4 p}{(2 \pi)^4} \rho_{\chi}^2 n_B(p_0) ( 1 + n_B(p_0)) \,, \label{dissnum}
\ee
where $n_{B}=(e^{p_0/T}-1)^{-1}$ is the Bose-Einstein distribution function, and $\rho_{\chi}=\rho_{\chi_i}$ the scalar spectral function: 
\be
\rho_\chi= \frac{4 \omega_p \Gamma_\chi}{(p_0^2 -\omega_p^2)^2+4 \omega_p^2 \Gamma_\chi^2}\,,
\ee
where $\omega_p^2=p^2 + m_\chi^2$, and $\Gamma_\chi$ is the thermal decay width of the scalars into a pair of massless fermions:
\be
\Gamma_\chi = \frac{h^2}{16 \pi} \frac{m_\chi^2}{\omega_p} \left[ 1 + \frac{2 T}{p} \ln \frac{1+e^{-\omega_+/T}}{1+ e^{-\omega_-/T}} \right]\,, 
\ee
with $\omega_{\pm}= \omega_p \pm p$. In an expanding universe, this expression holds for $T/H > 1$. 

\begin{figure}[t]
  \centering
  \includegraphics[scale=0.35]{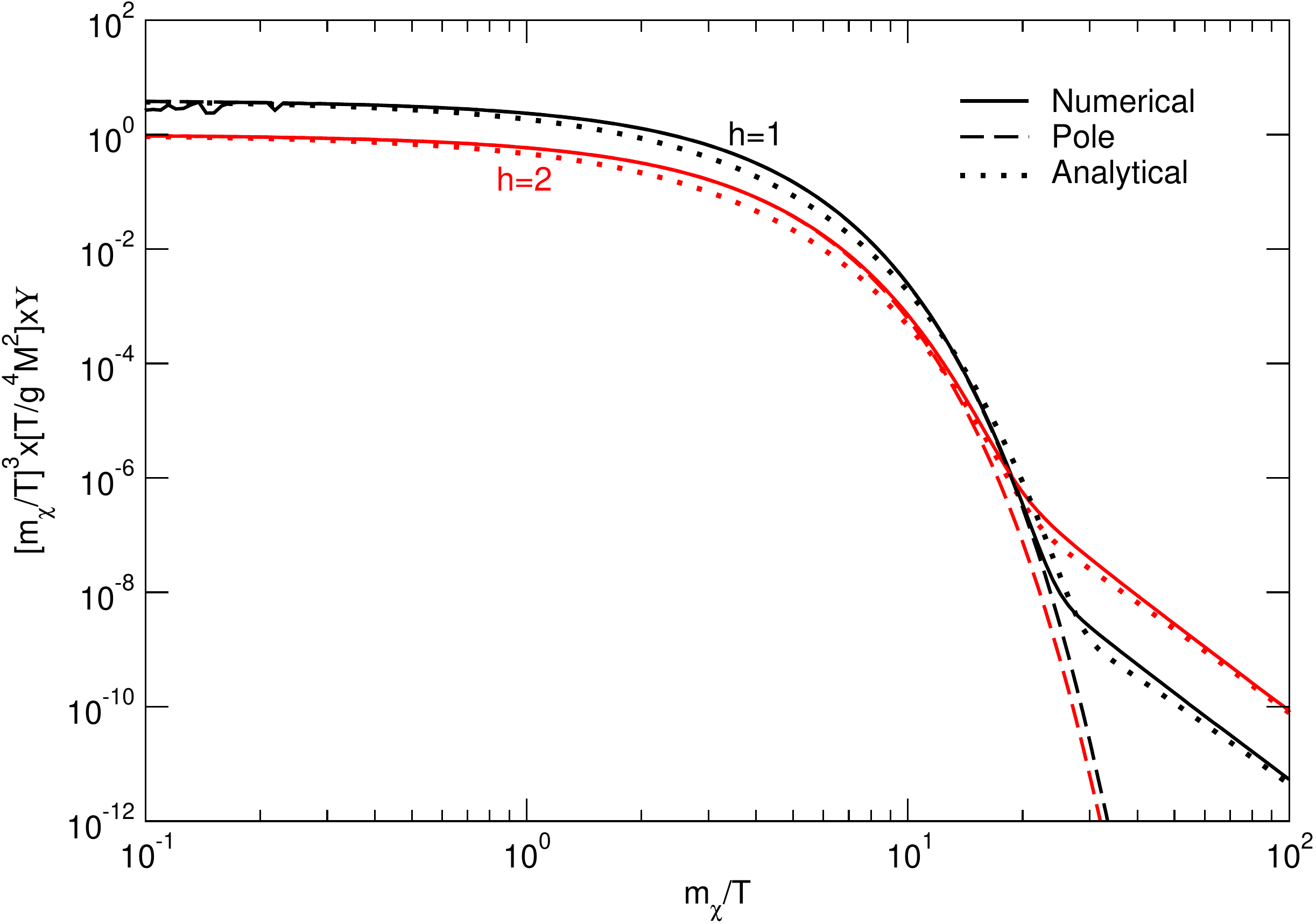} 
  \caption{Dissipative coefficient, modulus the prefactor $(g^4M^2/T)(T/m_\chi)^3$, versus the ratio $m_\chi/T$, for different values of the Yukawa coupling $h$ as indicated in the plot. We compare the numerical integration from Eq. \eqref{dissnum} (solid line), with the pole approximation Eq. \eqref{disspole} (dashed line) and the approximation used in this work Eq. \eqref{dissapprox} (dotted line).  } 
         \label{plotdiss}
\end{figure}

When $m_\chi \ll T$, in the high $T$ regime, the dissipative coefficient can be computed analytically expanding around the pole, $p_0 = \omega_p \pm i  \Gamma_\chi$ \cite{SWLI},
\be
\Upsilon_P \simeq \frac{4 g^4 M^2}{T} \int \frac{d^3 p}{(2 \pi)^3} \frac{n_B(p_0) ( 1 + n_B(p_0))}{\Gamma_\chi \omega_p^2} \simeq 
\frac{4 g^4 M^2}{h^2} \cdot \frac{T^2}{m_\chi^3} \left( 1 + \frac{1}{\sqrt{2 \pi}} \left(\frac{m_\chi}{T}\right)^{3/2} \right) e^{-m_\chi/T} \,. \label{disspole} 
\ee
However this approximation fails in the opposite limit, the low-$T$ regime when $m_\chi/T \gg 1$ and then $\Upsilon \propto T^7$. A better approximation, obtained directly from a fitting to the numerical integration, is given by:
\be
\Upsilon = \frac{4 g^2}{h^2} \cdot \frac{g^2 M^2}{T} \left(\frac{T}{m_\chi}\right)^3  \left( e^{-0.77 m_\chi/T} + 0.0135 h^6 e^{-20 T/m_\chi} \left( \frac{T}{m_\chi}\right)^5 \right) \,.  \label{dissapprox}
\ee
In Fig. (\ref{plotdiss}) we have compared the dissipative coefficient obtained from the numerical integration Eq. \eqref{dissnum}, the pole approximation Eq. \eqref{disspole}, and the approximation used in this work Eq. \eqref{dissapprox}. The pole approximation works very well even upto $m_\chi/T \sim 20$, but it only gives the exponential suppression for large masses and it does not reproduced the power-law contribution when $m_\chi/T \gg 1$. 

\section{Fluctuations in WI and the growing mode function}
\label{growing}

In order to get the amplitude of scalar primordial spectrum on superhorizon scales, we work with the comoving curvature perturbation at linear order ${ \cal R}$. In WI we have a mixture of two fluids, radiation and inflaton field,  and for each one we can define
\be
{\cal R}_\alpha= - \frac{H}{h_\alpha} \Psi_\alpha^{GI} \,,
\ee
with $h_\alpha=\rho_\alpha + p_\alpha$, $\rho_\alpha$, $p_\alpha$ being the energy density and pressure of the background fluid, and $\Psi_{\alpha}^{GI}$ the gauge invariant (GI) momentum perturbation. The total comoving curvature perturbation is given by
\be
   {\cal R}= \frac{h_\phi}{h_\phi + h_r} {\cal R}_\phi + \frac{h_r}{h_\phi + h_r} {\cal R}_r \,. \label{Rdef}
\ee
During slow-roll, one can  show that the momentum perturbations fulfil the relation \cite{growing2,reexamination}: 
\be
\Psi_r^{GI} \simeq Q \Psi_\phi^{GI} \,,
\ee
on super-horizon scales, and using the slow-roll Eq. \eqref{rhorsl} we simply have:
\be
{\cal R} \simeq - \frac{(1+ Q) H }{h_r + h_\phi} \Psi_\phi^{GI} \simeq {\cal R_\phi} \,,
\ee
which means that the primordial perturbations are adiabatic and there are no entropy perturbations. The GI inflaton momentum perturbation is given by $\Psi_\phi^{GI} = - \dot \phi \delta \phi^{GI}$, $\delta \phi^{GI}$ being the GI inflaton fluctuation, and using $h_\phi= \dot \phi^2$ we have 
\be
   {\cal R}_\phi = \frac{H}{\dot \phi} \delta \phi^{GI} \,.
\ee
The spectrum is then given by: 
\be
P_{\cal R}(k) = \frac{k^3}{ 2 \pi^2} |{\cal R}_k|^2 \simeq \left( \frac{H}{\dot \phi^2} \right)^2 P_{\phi} (k) \,,
\ee
where ${\cal R}_k$ is the perturbation in Fourier space, $P_{\phi}(k)$ the inflaton spectrum.

Gauge invariant perturbations at first order are built combining field, momentum and energy density perturbations with metric ones. At linear order, the FLRW metric is given by:
\be
ds^2= - (1 + 2 \alpha) dt^2 -2 a \partial_i \beta dx^i dt + a^2(t)[ \delta_{ij} ( 1 + 2 \varphi)+2 \partial_i \partial_j \gamma] dx^i dx^j \,.
\ee
We work in the longitudinal, shear-free gauge, in which $\chi=a(\beta + a \dot \gamma)=0$ and $\alpha=-\varphi$. The gauge invariant momentum, field and radiation energy density  perturbations at linear order are given by:
\bea
\Psi^{GI} &=& \Psi - (\rho + p) \varphi/H \,, \\
\delta \phi^{GI} &=& \delta \phi - \frac{\dot \phi}{H} \varphi \,, \\
\delta \delta \rho_r^{GI} &=& \delta \rho_r - \frac{\dot \rho_r}{H} \varphi \,.
\eea
Instead of writting directly with the evolution equations for the GI perturbations, we found numerically more convenient to work with the perturbed Eqs. for field and radiation energy density, including the metric perturbations, and obtain the GI curvature perturbation from Eq. \eqref{Rdef}. The equations for the coupled system of inflaton, radiation and metric perturbations can be found in \cite{growing2, growi<ng3}. Metric perturbations in the longitudinal gauge are given by combining the Einstein Eqs. at linear order: 
\bea
(H \alpha - \dot \varphi) &=& - \frac{1}{2 m_P^2} \Psi \,, \\
-\frac{k^2}{a^2} \varphi + 3H (H \alpha -\dot \varphi)  &=& -\frac{1}{2m_P^2} \delta \rho \,,
\eea
with $\alpha = -\varphi$, where $\delta \rho$ ($\delta \Psi$)  is the total energy (momentum) density perturbation. The Eqs. for the radiation fluctuations are given by (in Fourier space):
\bea
\delta \dot \rho_r + 4 H \delta \rho_r &=& -3 H (1 + w_r) \rho_r \dot \varphi + \frac{k^2}{a^2} \Psi_r + \delta Q_r + Q_r \alpha \,, \label{deltarhor}\\
\dot \Psi_r + 3 H \Psi_r &=& - w_r \delta \rho_r - (1+ w_r) \rho_r \alpha - \Upsilon \dot \phi \delta \phi \,, \label{psir}
\eea
where $w_r=1/r$, $Q_r= \Upsilon \dot \phi^2$ is the source term for the radiation that we have in Eq. \eqref{rhor}, and $\delta Q_r$ its perturbation 
\be
\delta Q_r = \delta \Upsilon \dot \phi^2 + 2 \Upsilon \dot \phi \delta \dot \phi - 2 \alpha \Upsilon \dot \phi^2 \,.
\ee
Finally, the Eq. for the inflaton field perturbation is given by a Langevin-like equation, including the stochastic Gaussian noises, quantum $\xi_q$ and thermal $\xi_T$:
\be
\delta \ddot \phi + 3 H \delta \dot \phi + \left( \frac{k^2}{a^2} + V_{\phi \phi} \right) \delta \phi = \sqrt{2} H \xi_q + \sqrt{2 \Upsilon T}\xi_T - \delta \Upsilon \dot \phi 
+ \dot \phi (3 (H \alpha - \dot \varphi) + \dot \alpha) + (2 \ddot \phi + 3 H \dot \phi) \alpha - \Upsilon ( \delta \phi - \alpha \dot \phi) \,, \label{deltaphi}
\ee
where $\langle \xi_\alpha (k, t) \xi_\alpha (k', t') \rangle= \delta^{(3)}(k -k') \delta (t-t')$. In order to get the comoving curvature spectrum we have to integrate the system of Eqs. \eqref{deltarhor}, \eqref{psir}, \eqref{deltaphi} for different realisations of the noise, and take the average:
\be
P_{\cal R} = \frac{k^3}{2 \pi^2} \langle | {\cal R}|^2 \rangle_\xi \,.
\ee
\begin{figure}[t]
  \centering
  \begin{tabular}{ccc}
        \includegraphics[scale=0.35]{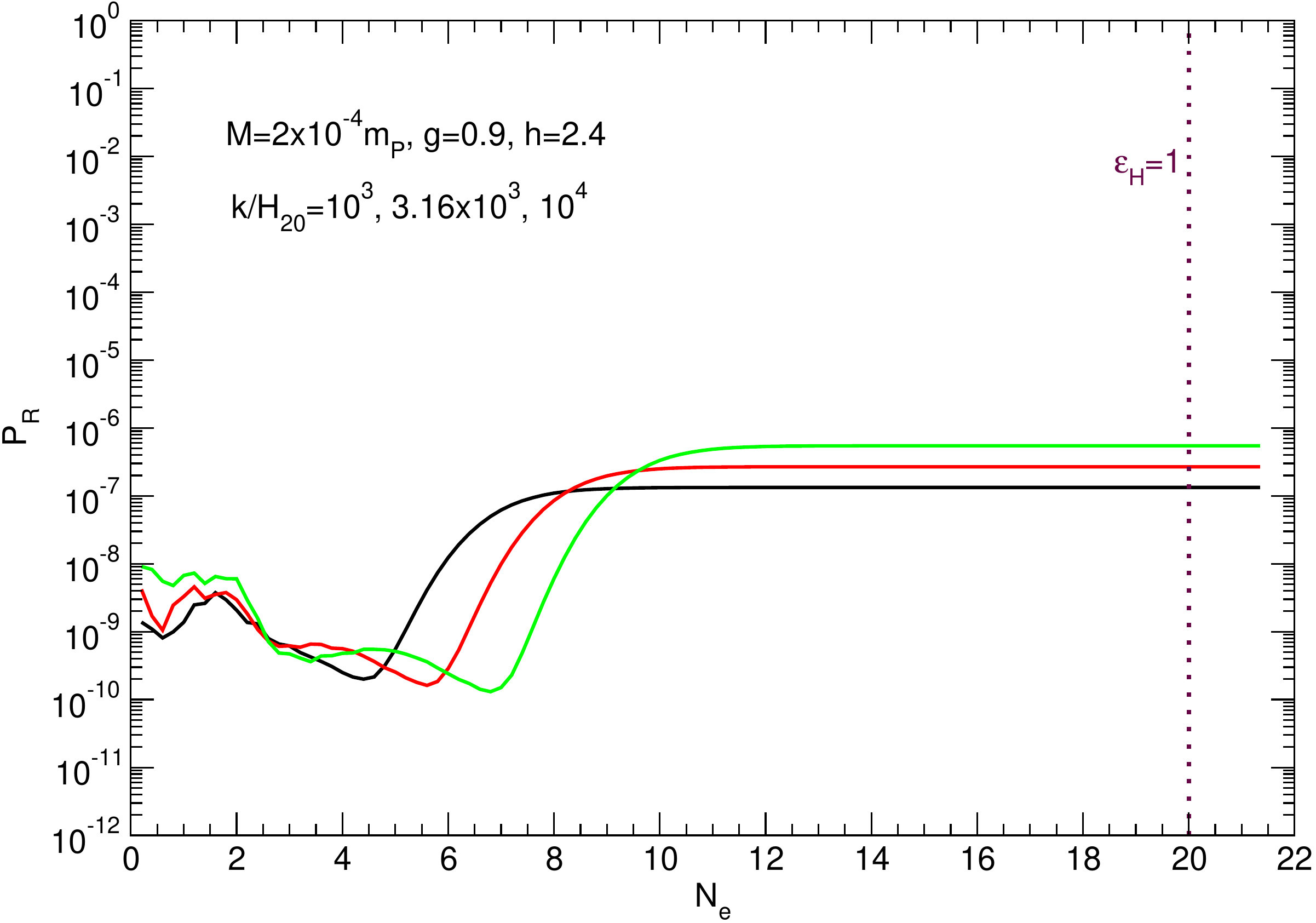} && 
	\includegraphics[scale=0.35]{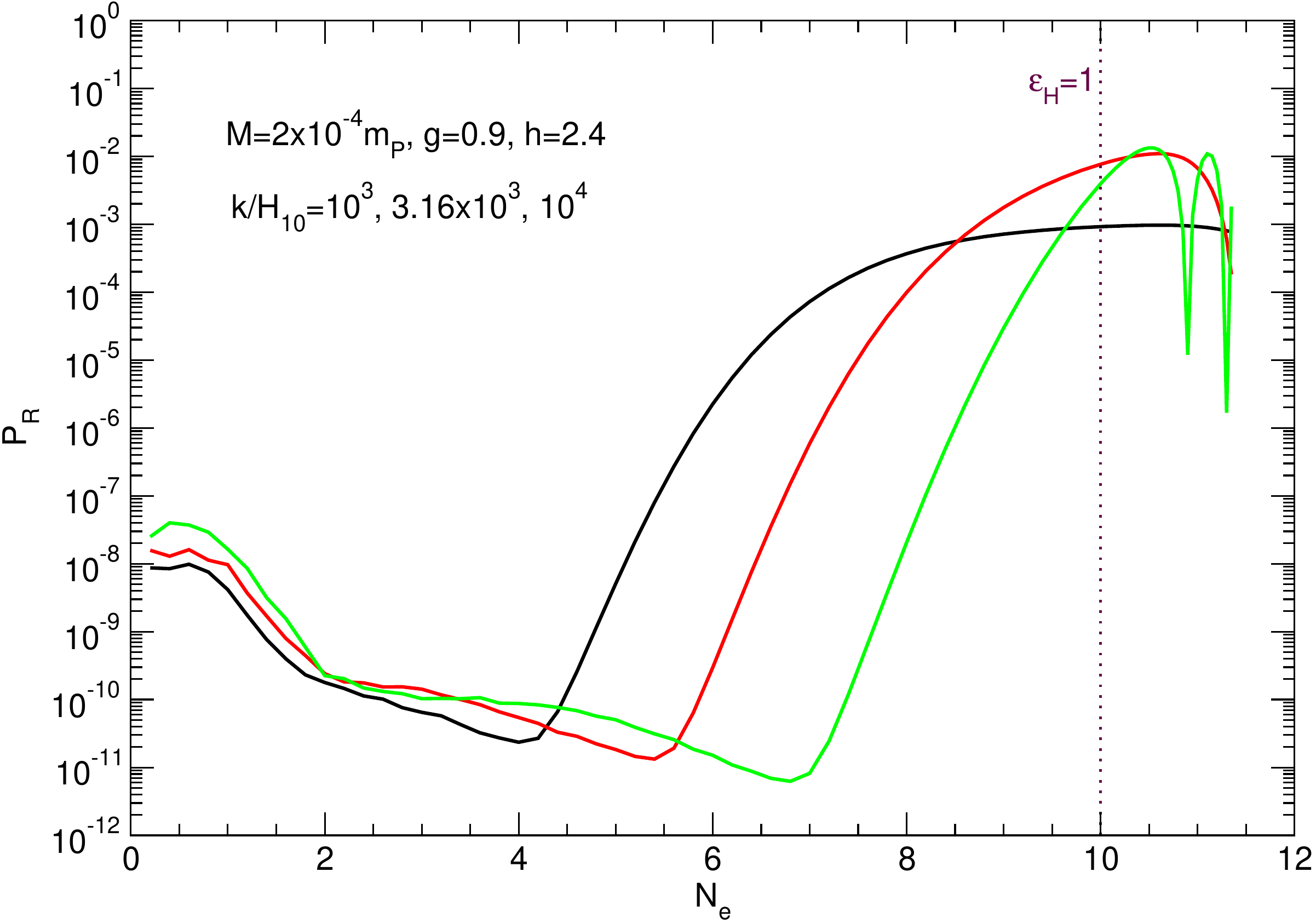} 
            \end{tabular}
  \caption{Numerical evolution of the  $P_{\cal R}$ for different mode values. On the LHS we start the integration 20 e-folds before the end of inflation (the line labelled $\epsilon_H=1$), for the set of comoving wavenumbers given in units of the initial Hubble parameter. We do the same on the RHS but starting at 10 e-folds before the end. Other parameter values as indicated in the plot. On the LHS we have initially $Q_{20}=28$ and $\kappa_{20}=0.5$, while on the RHS we have 
    $Q_{20}=50$ and $\kappa_{20}=1.5$.}
  \label{plotPRNe}
\end{figure}
Notice that the Eqs. are coupled through the terms proportional to
\be
\delta \Upsilon (T, \phi) =
\frac{d \ln \Upsilon}{d \ln T} \frac{\delta T}{T}  + \frac{d \ln \Upsilon}{d \ln \phi} \frac{\delta \phi}{\phi} \,,
\ee
for a general dissipative coefficient $\Upsilon(T,\phi)$. When $\kappa= d \ln \Upsilon/d \ln T  > 0$ this leads to the amplification of the radiation energy density perturbation and therefore to that of the field fluctuation \cite{growing1}. For the dissipative coefficient and the inflationary model considered in this work, an example of the numerical evolution of $P_{\cal R}$ is given in Fig. \eqref{plotPRNe}. On the LHS we have chosen the initial background values for field and radiation energy density such that we start the integration of the perturbations 20 e-folds before the end, while on the RHS we start at 10 e-folds.  We choose the comoving wavenumber of the perturbations in units of the initial Hubble parameter in each case. Initial field fluctuations are taken to be in vacuum, while we set the radiation to zero. We have checked that the evolution does not depend on the choice of initial conditions for the perturbations because of the stochastic nature of the system: in particular the thermal noise term will bring them quickly into their thermal values. Perturbations are amplified before they become superhorizon when $k< aH$, and afterwards the comoving curvature perturbation freezes-out. Due to the increasing behavior of both $Q$ and mainly $\kappa$, the primordial curvature spectrum will peak at slightly larger modes than the last one crossing the horizon.

\begin{figure}[t]
  \centering
  \begin{tabular}{ccc}
        \includegraphics[scale=0.35]{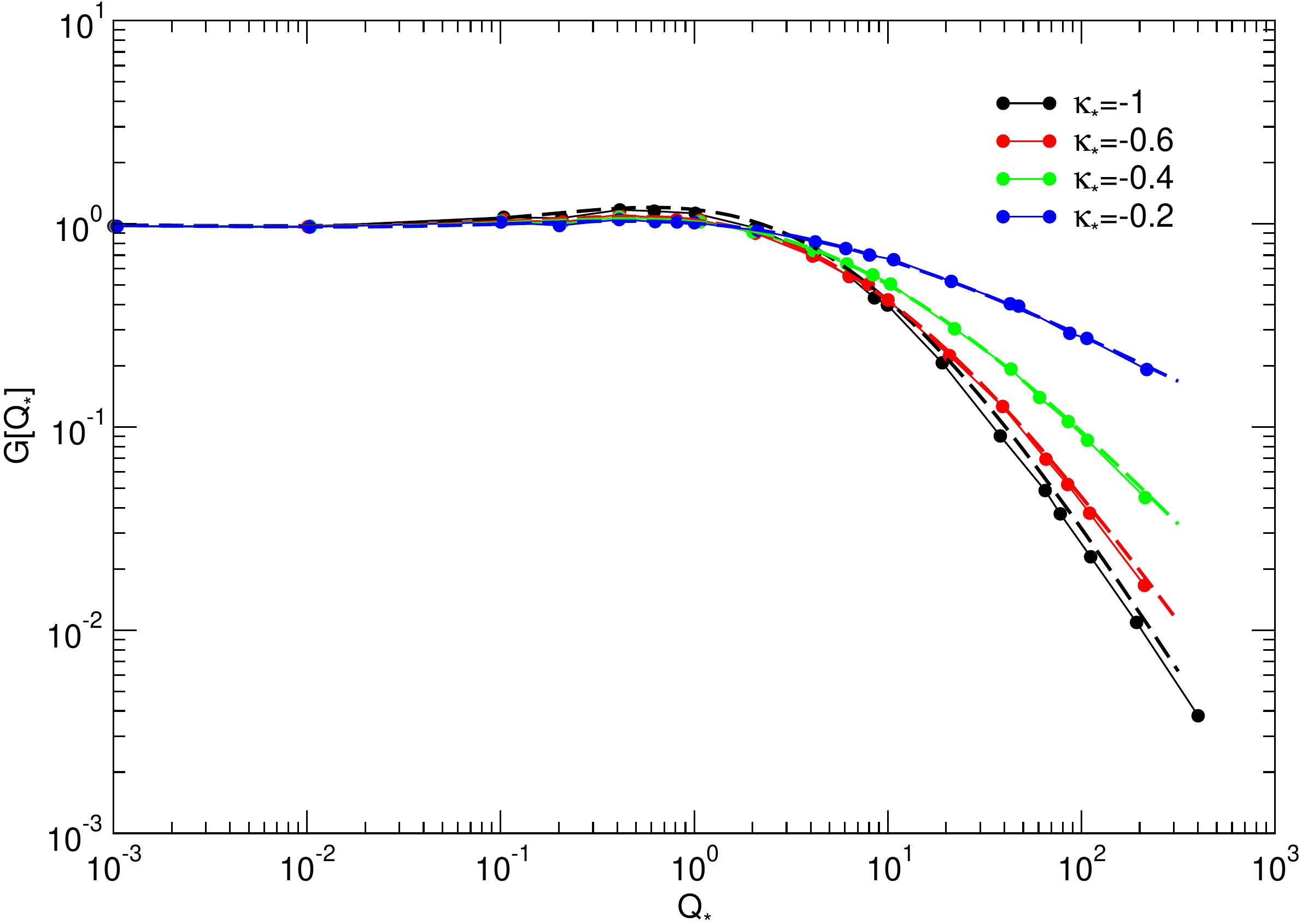} && 
	\includegraphics[scale=0.35]{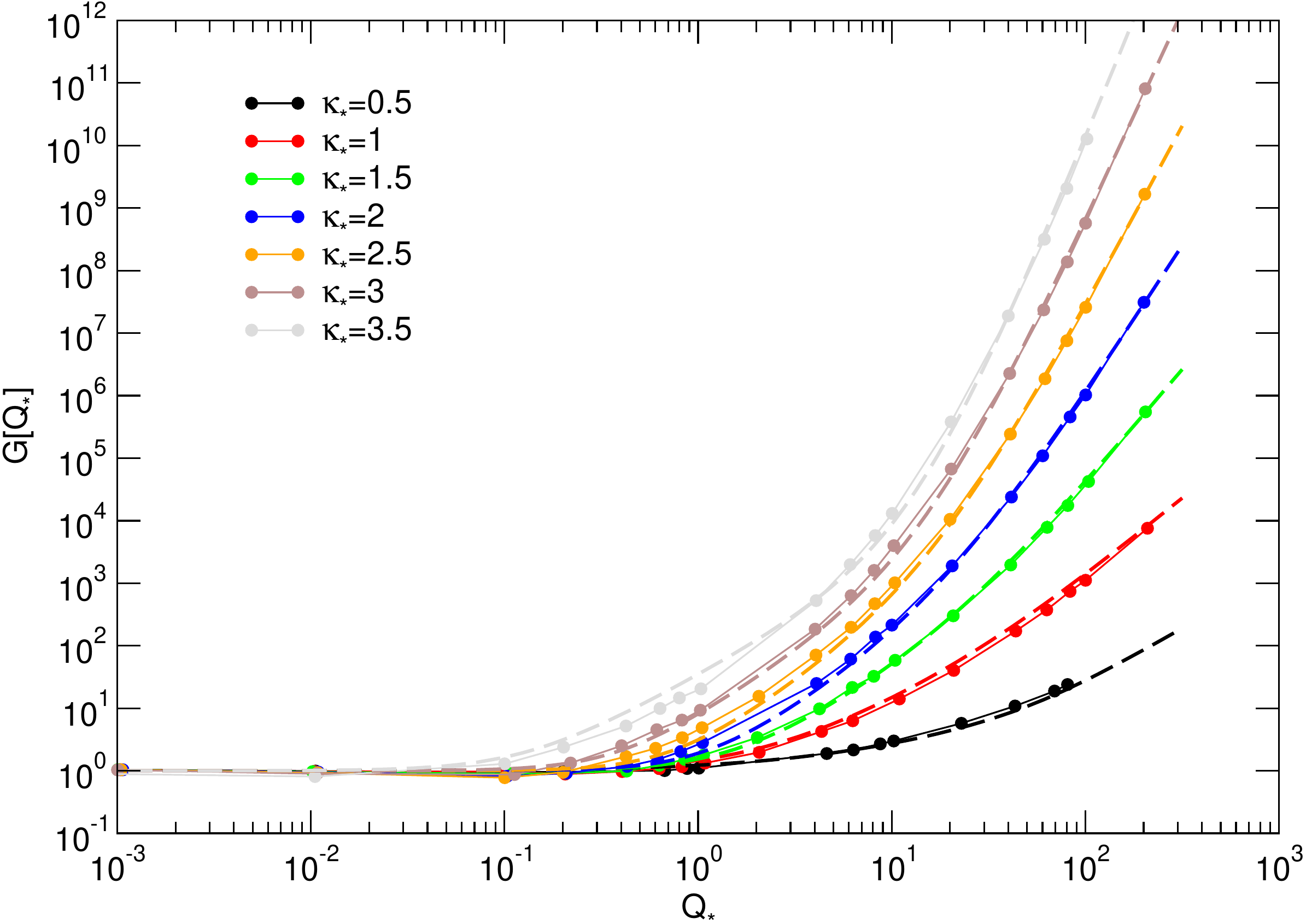} 
            \end{tabular}
	\caption{Amplitude of the primordial spectrum $P_{{\cal R}_*}$ obtained numerically for different values of $Q_*$ and $\kappa_*$, normalised by the analytical value $P_{{\cal R}_*}^{{\rm nogr}}$  valid for $\kappa=0$. Numerical values are indicated by filled circles, dashed lines are the fitted functions $G[Q_*]$ given in Eqs. \eqref{GQposapp} for $\kappa_*>0$ (RHS), and \eqref{GQnegapp} for $\kappa_*<0$ (LHS). } \label{plotgrowing}
\end{figure}

For constant dissipation with neither $T$ or $\phi$ dependence, the analytical solution was obtained in \cite{rudneisilva}, and the amplitude of the comoving curvature perturbation is given by:
\be
P_{{\cal R}_*}^{\rm nogr} = \left( \frac{H_*^2}{2 \pi \dot \phi_*} \right)^2 \left( 1 + 2 n_* + \frac{T_*}{H_*}\cdot \frac{2 \sqrt{3} \pi Q_*}{\sqrt{3 + 4 \pi Q_*}}\right)  \,. \label{PRnogrowing}
\ee
where the label ``nogr'' means without growing mode. Otherwise the system of Eqs. have to be integrated numerically in order to get the modification to the above expression, i.e., the growing mode function $G[Q_*]$ given in \eqref{PRstar}. We have done this for different values of constant $\kappa$ in $\Upsilon= C_\Upsilon T^\kappa$, and a quartic chaotic potential. By varying the value of $C_\Upsilon$ we can tune the value of $Q$ at horizon crossing.
The results of comparing our numerical results for $P_{{\cal R}_*}$ with the analytical expression Eq. \eqref{PRnogrowing}, which gives the function $G[Q, \kappa]$, are shown in Fig. \eqref{plotgrowing}. We have parametrized this function as:
\bea
G[Q,\kappa] &=& (1 +  e^{\alpha_s} Q^{\beta_s} + e^{\alpha_w} Q^{\beta_w})^\kappa \,, \;\;\; \kappa > 0 \,, \label{GQposapp}\\
G[Q, \kappa] &=& \frac{(1 + a_0 Q^{a_1})^{a_5}}{(1 + a_2 Q^{a_3})^{a_4}} \,, \;\;\; \kappa \leq 0 \label{GQnegapp}\,,
\eea
which generalized the standard parametrizations found in the literature for the function $G[Q_*]$ \cite{growing1,chaoticwarm,growing2}. By comparing with the numerical results we obtain the coefficients: 
\bea
\alpha_w &=&-1.486 + 0.7091\kappa  \,, \label{alw}\\
\beta_w  &=& 1.711 - 0.3499\kappa    \,,\label{betw}\\
\alpha_s &=& -5.168 + 0.5105\kappa  \,, \label{als}\\
\beta_s  &=& 2.692 - 0.1472\kappa   \,, \label{bets}
\eea
when $\kappa >0$, and:
\bea
 a_0 &=& 18.547 + 171.01  \kappa + 765.78  \kappa^2 + 1758.7  \kappa^3+ 2158.7  \kappa^4 + 1338.8  \kappa^5 + 328.71  \kappa^6 \,, \label{a0}\\
 a_1&=& 1.1094 + 4.619  \kappa + 21.647  \kappa^2 + 50.429  \kappa^3+ 61.783  \kappa^4 + 37.797  \kappa^5 + 9.098  \kappa^6 \label{a1} \,, \\
 a_2 &=& 4.0275 + 28.686  \kappa + 103.74  \kappa^2 + 199.28  \kappa^3+ 210.14  \kappa^4 + 114.8  \kappa^5 + 25.452  \kappa^6 \,, \label{a2} \\
 a_3 &=& 0.68259 + 0.99858  \kappa + 3.1507  \kappa^2 + 3.7451  \kappa^3+ 0.563  \kappa^4 - 1.9247  \kappa^5 - 0.95376  \kappa^6 \,, \label{a3}\\
 a_4 &=&\kappa ( -9.9143 + 34.481  \kappa + 250.17  \kappa^2 + 662.93  \kappa^3+ 880.09  \kappa^4 + 571.9  \kappa^5 + 144.43  \kappa^6) \,, \label{a4}\\
 a_5&=& \kappa ( -4.3323 + 36.915  \kappa + 284.89  \kappa^2 + 803.87  \kappa^3+ 1104.5  \kappa^4 + 734.8  \kappa^5 + 188.95  \kappa^6) \,, \label{a5}
 \eea
 when $\kappa <0$. 

\section{Parameter values and primordial spectrum at the end of inflation}
\label{param}

Although our fitting functions for  the growing mode $G[Q_*]$ work well when $\kappa \simeq$Constant, we have checked numerically that the approximation fails at the end of inflation when $\kappa$ becomes positive and starts varying. Therefore, to compute the primordial spectrum upto to the end, using the Eqs. \eqref{deltarhor}, \eqref{psir}, \eqref{deltaphi}, we have first selected a set of parameter values $M$, $g$ and $h$ that fulfilled our requirements: the predicted spectral index and tensor-to-scalar ratio are consistent with observations, while around 10-20 efolds before the end we have the transition towards the SDR with $\kappa > 0$ and the amplification of the spectrum. The list of the chosen parameter values, including the value of the dissipative ratio $Q_*$ and the comoving wavenumber for the last possible mode exiting the horizon $k_{\rm end}$ are given in Table \ref{tablekend}.

\begin{table}[h]
  \begin{tabular}{lclclclcl}
    $M/m_P$ & &$g$ && $h$ && $Q_*$ && $k_{\rm end}$[Mpc$^{-1}$] \\
    \hline \\
 1$\times 10^{-4}$ && 1.4 && 3.6 &&  0.04 &&   2.140545$\times 10^{20}$ \\ 
 2$\times 10^{-4}$ && 0.9 && 2.4 &&  0.71 &&   3.026596$\times 10^{20}$ \\ 
 3$\times 10^{-4}$ && 0.5 && 1.6 &&  0.18 &&   2.713322$\times 10^{20}$ \\ 
 4$\times 10^{-4}$ && 0.4 && 1.2 &&  0.55 &&   2.806728$\times 10^{20}$ \\ 
 5$\times 10^{-4}$ && 0.3 && 1.0 &&  0.22 &&   2.774857$\times 10^{20}$ \\ 
 6$\times 10^{-4}$ && 0.3 && 1.1 &&  0.61 &&   3.618087$\times 10^{20}$ \\ 
 7$\times 10^{-4}$ && 0.2 && 0.7 &&  0.13 &&   2.636189$\times 10^{20}$ \\ 
 \hline
  \end{tabular}
  \caption{Values of $k_{\rm end}=(a H)_{\rm end}$ for different model parameters.}
\label{tablekend}
\end{table}

In order to compute the induced GW spectrum by the scalar perturbations, once the primordial curvature spectrum was obtained, we have fitted the numerical values using the function:
\be
\ln P_{\cal R} [k] = c_0 + (1 - {\rm exp}(b_0 x -b_1)) P_5[x] \,,
\ee
where $x=k/k_{\rm end}$, $P_5[x] = \sum_{i=0,5} a_i x^i$, and $c_0=-25$. The values of the coefficients $a_i$ and $b_i$ can be found in Table \ref{tableparamaibi}. This fitting function works for $k/k_{\rm end} > 10^{-12}$.

\begin{table}[h]
  \begin{tabular}{lllllllll}
    $M/m_P$ & $a_0$ & $a_1$& $a_2$ & $a_3$& $a_4$ & $a_5$ & $b_0$ & $b_1$\\
    \hline \\
 1$\times 10^{-4}$ &  10.101 &   2.7069 &   0.32689 &   0.018567 &   0.00051652 &   5.6499$\times 10^{-6}$  & 1.12985 &    3.7934  \\ 
 2$\times 10^{-4}$ &  18.351 &   1.7829 &   0.14145 &   0.0083661 &  0.00027298 &   3.45$\times 10^{-6}$    & 1.81186 &    5.7949  \\ 
 3$\times 10^{-4}$ &  15.487 &   1.94 &     0.14802 &   0.007783 &   0.00025559 &   3.4977$\times 10^{-6}$  & 1.53848 &    5.09359 \\ 
 4$\times 10^{-4}$ &  18.233 &   1.8019 &   0.14391 &   0.0086305 &  0.00028623 &   3.6692$\times 10^{-6}$  & 1.60554 &    5.32719 \\ 
 5$\times 10^{-4}$ &  15.963 &   1.8733 &   0.14483 &   0.0080364 &  0.00026916 &   3.6427$\times 10^{-6}$  & 1.54388 &    5.10092 \\ 
 6$\times 10^{-4}$ &  15.25 &    1.5821 &   0.12109 &   0.0066757 &  0.00021502 &   2.7598$\times 10^{-6}$  & 1.56107 &    4.99287 \\ 
 7$\times 10^{-4}$ &  14.745 &   1.944 &    0.14338 &   0.0067079 &  0.00020095 &   2.6645$\times 10^{-6}$  & 1.51569 &    5.01758 \\ 
 \hline
  \end{tabular}
  \caption{Values of $a_i$, $b_i$  for different model parameters. We only quote the value of $M$ for each set of parameters.}
\label{tableparamaibi}
\end{table}

\acknowledgments

This work has been partially supported by MICINN (PID2019-105943GB-I00/AEI/10.13039/501100011033) and ``Junta de Andaluc\'ia" grants  P18-FR-4314  and  A-FQM-211-UGR18.


\begin{thebibliography}{99}
\bibitem{growing3} A nice summary of the Eqs. for the perturbations in WI can be found in Appendix B of Ref. \cite{rudneidas}.
\end{thebibliography}
\end{document}